\documentclass{aa}  

\usepackage{graphicx}
\usepackage{txfonts}
\usepackage{natbib}
\usepackage{subfig}
\usepackage{float}
\usepackage{xcolor}
\usepackage[flushleft]{threeparttable}
\bibpunct{(}{)}{;}{a}{}{,} 
\usepackage[normalem]{ulem} 

\newcommand{\modif}[1]{\textnormal{#1}}

\newcommand{\modiflet}[1]{\textnormal{#1}}

\begin{document}

   \title{What does a typical full disc around a post-AGB binary look like?\thanks{Based on observations collected at the European Southern Observatory under ESO programmes 094.D-0865, 0102.D-0760, 60.A-9275, and 0104.D-0739}}
   
   \subtitle{Radiative transfer models reproducing PIONIER, GRAVITY, and MATISSE data}

   \author{A. Corporaal\inst{1}
           \and J. Kluska\inst{1}
           \and H. Van Winckel\inst{1}
           \and D. Kamath\inst{2,3,4}
           \and M. Min\inst{5}}

   \institute{Institute of Astronomy, KU Leuven,
              Celestijnenlaan 200D, 3001 Leuven, Belgium\\
             \email{akke.corporaal@kuleuven.be}
             \and School of Mathematical and Physical Sciences, Macquarie University, Sydney, NSW, Australia
    \and Astronomy, Astrophysics and Astrophotonics Research Centre, Macquarie University, Sydney, NSW, Australia
    \and INAF, Observatory of Rome, Via Frascati 33, 00077 Monte Porzio Catone (RM), Italy
    \and SRON Netherlands Institute for Space Research, Niels Bohrweg 4, 2333 CA, Leiden, The Netherlands}
   \date{Received ; accepted}
 
  \abstract
   {
   Stable circumbinary discs around evolved post-asymptotic giant branch (post-AGB) binary systems composed of gas and dust show many similarities with protoplanetary discs around young stellar objects. These discs can provide constraints on both binary evolution and the formation of macrostructures within circumstellar discs. 
   Here we focus on one post-AGB binary system: IRAS\,08544-4431.}
   {We aim to refine the physical model of IRAS\,08544-4431 with a radiative transfer treatment and continue the near-infrared and mid-infrared interferometric analysis covering the $H$, $K$, $L$, and $N$ bands.
   Results from geometric modelling of these data in our previous study constrain the shape of the inner rim of the disc and its radial dust structure.
   We aim to capture the previously detected amount of over-resolved flux and the radial intensity profile at and beyond the inner dust disc rim to put constraints on the physical processes in the inner disc regions.}
   {We used a three-dimensional Monte Carlo radiative transfer code to investigate the physical structure of the disc by reproducing both the photometry and the multi-wavelength infrared interferometric dataset.
   We first performed a parametric study to explore the effect of the individual parameters and selected the most important parameters, which were then used in a thorough grid search to fit the structural characteristics.
   We developed a strategy to identify the models that were best able to reproduce our extensive multi-wavelength dataset.}
   {We find a family of models
   that successfully fit the infrared photometric and interferometric data in all bands.
   These models show a flaring geometry with efficient settling.
   Larger grains are present in the inner disc as probed by our infrared interferometric observations.
   Some over-resolved flux component was recovered in all bands, but the optimised models still fall short in explaining all the over-resolved flux.
   This suggests that another dusty structure within the system that is not included in our models plays a role. 
   The structure of this over-resolved component is unclear, but it has a colour temperature between 1400 and 3600\,K.
   }
   {Multi-wavelength infrared interferometric observations of circumstellar discs allow the inner disc regions to be studied in unprecedented detail. 
   The refined physical models can reproduce most of the investigated features, including the photometric characteristics, the radial extent, and the overall shape of the visibility curves.
   Our multi-wavelength interferometric observations combined with photometry show that the disc around IRAS\,08544-4431 is similar to protoplanetary discs around young stars with similar dust masses and efficient dust growth.
   The resulting disc geometry is capable of reproducing part of the over-resolved flux, but to fully reproduce the over-resolved flux component, an additional component is needed.
   Multi-scale high-angular-resolution analysis combining VLTI, VLT/SPHERE, and ALMA data is needed to fully define the structure of the system.}

   \keywords{Stars: AGB and post-AGB - techniques: interferometric - binaries: general - protoplanetary discs - circumstellar matter}

   \maketitle
%
\section{Introduction}
Circumstellar discs composed of gas and dust are found at various stellar evolutionary stages.
One class of evolved stellar systems that show such circumstellar discs are post-asymptotic giant branch (post-AGB) binaries.
Their spectral energy distributions (SEDs) show infrared (IR) excesses, pointing towards the presence of hot dust.
It has now been well established that this property of the SEDs provides observational evidence for the presence of stable, massive circumbinary discs around these systems \citep[e.g.][]{deRuyter_2006, vanwinckel_2003, vanWinckel_2017, Kamath_2014, Kamath_2015,Kluska_2022}.
Observational properties of these systems were recently reviewed by \citet{VanWinckel_2018}.

These discs show evidence of stability, and a Keplerian velocity field has been spatially resolved at millimetre wavelengths in CO in several objects \citep{Bujarrabal_2013, Bujarrabal_2015, Bujarrabal_2017, Bujarrabal_2018, CallardoCava_2021}. More objects were detected in single-dish observations, and their narrow CO profile is also indicative of rotation \citep{Bujarrabal_2013a}.
Moreover, the dust grains show evidence of strong processing in a stable environment that results in grain growth and a high degree of crystallinity, as revealed by IR spectroscopic observations and millimetre photometry \citep[e.g.][]{Gielen_2011, Sahai_2011}.

While the presence of these circumbinary discs around post-AGB binaries is well established, their formation, structure, and evolution are still insufficiently understood. 
There is observational evidence for interactions between the system's components. 
Indirect observational evidence for disc--binary interactions is provided by analyses of the time series of spectroscopic observations and photospheric abundance determinations.
The former show fast outflows or jets originating from an accretion disc around the companion \citep{Gorlova_2012, Gorlova_2015, Bollen_2017, Bollen_2019, Bollen_2022}.
To launch the jets, the mass-accretion rates onto the companion are found to be of the order of $10^{-6}$ to $10^{-4}$ M$_\odot$/yr \citep{Bollen_2020}. 
Such a mass-accretion rate cannot be due to the mass-loss rate of the post-AGB primary and points towards accretion from the circumbinary disc as the main feeding mechanism of the circum-companion disc.
Additional evidence for accretion from the circumbinary disc comes from the depletion of refractory elements observed on the photosphere of the post-AGB primary itself \citep[e.g.][]{VanWinckel_1995, Maas_2005, Giridhar_2005, Oomen_2018}. 
Such a depletion is caused by the re-accretion of gas from the circumbinary disc \citep{Oomen_2019} while the refractory elements remain on the dust grains in the disc.
However, the cause of this dust--gas separation is still debated.

\citet{Kluska_2022} show a link between the strength of the photospheric depletion and the lack of near-IR excess in the SEDs of the systems (labelled transition discs).\ This link indicates that the most depleted targets are surrounded by discs with a large dust-free cavity.
In young stellar objects (YSOs) such a depletion pattern is also observed in targets that host a transition disc and is often linked to planet--disc interactions.


Constraining the physical processes in the inner disc regions and the disc--binary interactions of circumbinary discs around post-AGB binaries will improve our understanding of the late evolution of binary systems.
Interferometric techniques in the IR are needed to spatially resolve the inner disc regions as well as the inner binary. 

Two-dimensional radiative transfer modelling efforts of high-angular-resolution interferometric data of such discs \citep{Deroo_2006, Deroo_2007, Hillen_2014, Hillen_2015, Hillen_2017, Kluska_2018} have shown that passively irradiated disc models developed for protoplanetary discs are able to reproduce the data. 
This points towards a similar structure of discs around post-AGB binary systems and protoplanetary discs around YSOs. An important difference, however, is that the estimated lifetime of discs around post-AGB binaries is only of the order of ($10^4-10^5$\,yr), while protoplanetary discs live up to a few megayears. 

While there is growing evidence that initial phases of planet formation around YSOs can be short and that grain growth is very efficient on short timescales ($\sim 10^5$\,yr; e.g. \citealt{Sheehan_2018, SegurCox_2020, Cridland_2022, Lau_2022}), it is as yet unclear what the formation timescale of full-grown planets is. Discs around post-AGB binaries, thus, represent an interesting laboratory for testing processes for planet formation, and this in a different parameter space than around YSOs.



Here we focus on one post-AGB binary system, IRAS\,08544-4431, hereafter IRAS\,08544.
IRAS\,08544 is a luminous ($\sim 10500\,\mathrm{\,L_\odot}$) post-AGB star with a confirmed companion \citep{Maas_2003}.
The binary is surrounded by an optically thick circumbinary disc; its near-IR excess provides evidence for a full disc, with the inner rim located at the dust sublimation radius \citep{deRuyter_2006, Hillen_2016, Kluska_2018, Kluska_2022}.
It is classified as a Category 1 disc in \citet{Kluska_2022} and is the prototypical full disc surrounding a post-AGB binary.
Interferometric observations in the near-IR have revealed the structure of the inner rim of the disc via geometric models, image reconstruction techniques, and radiative transfer modelling \citep{Hillen_2016, Kluska_2018}.
With these techniques, a small resolved flux excess at the location of the companion \modiflet{has also been} identified.
As this excess cannot be explained by photospheric emission from the main sequence companion star, it likely originates from a circum-secondary accretion disc.

The radiative transfer model of IRAS\,08544 from \citet{Kluska_2018} provides a good fit to the SED and the $H$-band interferometric measurements.
These authors find that the disc inner rim coincides with the theoretical dust sublimation radius. 
Their disc model requires, however, an ad hoc over-resolved flux component of unknown origin on top of the disc model (see also Sects. \ref{section:strategy-extended_component} and \ref{section:Discussion_extended}).
By performing \modif{two-dimensional} geometric modelling of both near-IR and mid-IR interferometric observations, \citet{Corporaal_2021} reproduce the visibility data of the $H$, $K$, $L$, and $N$ bands and find that the inner rim of the circumbinary disc is rounded and puffed up.
In the near-IR, the inner rim, the stars, and a spatially extended component are detected. The over-resolved flux contribution to the total flux is relatively constant throughout the $H$, $K$, and $L$ bands. 
In the mid-IR, however, the stars contribute only a few per cent to the total flux, and thus the visibilities are mainly dominated by thermal emission and scattering from the circumbinary disc. 
While the \modif{two-dimensional geometric models (i.e. parameterised rings on the image plane)} are able to reproduce most features in the visibility data, we now want to \modif{develop a three-dimensional radiative
transfer model of a disc with self-consistent handling of dust settling} to infer the physical properties of the circumbinary disc.

Here we present such a physical model of the circumbinary disc around IRAS\,08544.
We aim to reproduce the main characteristics of both the observed photometry and the  interferometric visibilities in four IR bands.
We focus on investigating the over-resolved flux component and retrieve the radial profile of the inner disc regions using data of the three four-beam combiners at the Very Large Telescope Interferometer (VLTI): PIONIER, GRAVITY, and MATISSE
We summarise the observations in Sect.\,\ref{section:observations} and the physical setup of the radiative transfer  model in Sect.\,\ref{section:physicalsetup}.
We investigate our parameter space in Sect.\,\ref{section:parameterstudy} and use the results of the most impactful parameters to refine the disc model in Sect.\,\ref{section:application}.
We discuss the results and implications in Sect.\,\ref{section:discussion} and summarise our conclusions in Sect.\,\ref{section:conclusion}.

\section{Observations}
\label{section:observations}
 \subsection{Photometry}
The energetics of the target are taken from the catalogue of Galactic post-AGB binary systems of \citet{Kluska_2022} and we refer to this paper for a full description of the data collection.
In short, the full SED is assembled by collecting public broadband photometric data from a broad range of wavelengths (from 0.3\,$\mu$m to 0.8\,mm).
The parameters of the photospheric model of the post-AGB star were derived from Kurucz stellar atmosphere models \citep{Castelli_2003}.

 \subsection{Interferometry}
We used the IR interferometric dataset presented in \citet{Corporaal_2021}.
Here, we summarise the main characteristics of the data.
The dataset consists of observations obtained on the following three current four-telescope beam combiner instruments the VLTI at Mount Paranal in Chile: PIONIER \citep{LeBouquin2011}, GRAVITY \citep{GRAVITY2017}, and MATISSE \citep{Lopez2014,Lopez_2022}. 
These instruments provide simultaneous observations on six baselines and three independent closure phases per pointing.

The PIONIER instrument operates in the $H$ band (between 1.5\,$\mu$m and 1.85\,$\mu$m).
The dataset was taken in 2015 (prog. ID: 094.D-0865, PI: Hillen), using the three configurations on the 1.8\,m Auxiliary Telescopes (ATs). 
A log of these observations is reported in \citet{Hillen_2016}.

GRAVITY operates in the $K$ band (2.0 - 2.4\,$\mu$m).
The data were taken in 2018 and 2019 (prog. ID: 0102.D-0760, PI: Bollen) using the three configurations of the ATs at high resolution ($R\sim$ 4000) in single field mode.

The dataset taken with the MATISSE instrument covers the $L$ band (2.9-4.2\,$\mu$m) and $N$ band (8-13\,$\mu$m).
Observations were taken during 2019 (prog. ID 60.A-9275, PI: Kluska) and 2020 (prog. ID 0104.D-0739, PI: Kluska) with the three configurations of the ATs.
All observations \modiflet{were} taken in the low spectral resolution mode ($R \sim 30$).
$N$-band photometry was not taken during the 2019 observations such that the coherent flux measurements could not be normalised and the visibility amplitudes could not be determined.
As a result, the reported visibilities in the $N$ band are correlated fluxes.



\section{Physical setup}
\label{section:physicalsetup}
To infer the physical structure of the dusty circumbinary disc of IRAS\,08544 from the observations we used the Monte Carlo radiative transfer code MCMax3D \citep{Min_2009}.
In such a code, the photon packages emitted from the central stellar source are scattered, absorbed, or re-emitted by the dust particles. 
The user specifies the dust distribution by setting the disc structure and its density distribution as well as the dust grain properties and composition. 
In MCMax3D, the disc can be made of several zones with different parameters for the disc structure, which is defined by the dust disc inner and outer radii, the vertical dust settling, and the scale height.
This flexibility allows the exploration of complex disc geometries.

Vertical dust settling is handled self-consistently with a single parameter, the turbulent mixing strength, $\alpha$ \citep{Shakura_1973}, following the prescription of \citet{Mulders_2012}. 
Stronger turbulence mixing strengths imply more efficient settling as larger grains decouple from the gas.

The disc vertical scale height is defined by a power law:
\begin{equation}
    \label{Eq:scaleheight}
    h(r) =h_0 \left(\frac{r}{R_{\mathrm{in}}}\right)^\beta,
\end{equation}
where $h_0$ is the scale height at the radius we set to coincide with the disc inner radius, $R_\mathrm{in}$, and $\beta$ is the flaring exponent describing the disc curvature.
Grain sizes are distributed by a power law with index $q$:
\begin{equation}
    \label{Eq:dustgrains}
    n(a) \propto a^{-q}  \quad \mathrm{for} \quad a_\mathrm{min} < a < a_\mathrm{max}
,\end{equation}
where $a_\mathrm{min}$ and $a_\mathrm{max}$ are the minimum and maximum grain sizes, respectively.

The surface density profile is prescribed by a radial power law with index $p$ and radius $r$:
\begin{equation}
\label{Eq:surfacedensity}
    \Sigma(r)\propto\left(\frac{r}{r_c}\right)^{-p} 
.\end{equation}
The surface density is scaled to the dust mass in the disc.
We apply a two-zone model to adopt a double power law of the surface density to smooth the disc inner rim \citep{Hillen_2015,Kluska_2018}.
In our two-zone model, there is an additional parameter, the turn-over radius, $r_\mathrm{mid}$, at which the surface density profile (Eq. \ref{Eq:surfacedensity}) changes as we apply different values of $p$ in the zones. 
To ensure continuity in the full disc model, $r_\mathrm{mid}$ corresponds to the outer radius of the inner zone and the inner radius of the outer zone.
Besides the inner and outer radii of these discs and the surface density distribution, we do not consider differences in the inner and the outer disc.
The outcome of the Monte Carlo run is a three-dimensional temperature distribution within the disc. 
This thermal structure is used to calculate ray-traced synthetic spectra and images in a subsequent step.

\section{Parameter study}
\label{section:parameterstudy}

We first aim to understand the impact of the individual parameters on the SED and the visibilities. 
This will allow us to (1) select the parameters that impact the observables the most, such that we can constrain the disc parameters in a subsequent step to provide a good fit to the data, and (2) investigate the parameter space of the individual parameters. 
This section is organised as follows: we describe our reference model in Sect.\,\ref{section:referencemodel}, \modiflet{explain the over-resolved component in Sect.\,\ref{section:strategy-extended_component}}, discuss our strategy for our parametric study in Sect.\,\ref{section:strategy-parameter}, present the results in Sect.\,\ref{section:results-parameters}, and select the most promising parameters to meet the shortcomings of the reference model in Sect.\,\ref{section:selection-parameters}.

\begin{figure*}
\centering
  \includegraphics[width=17cm]{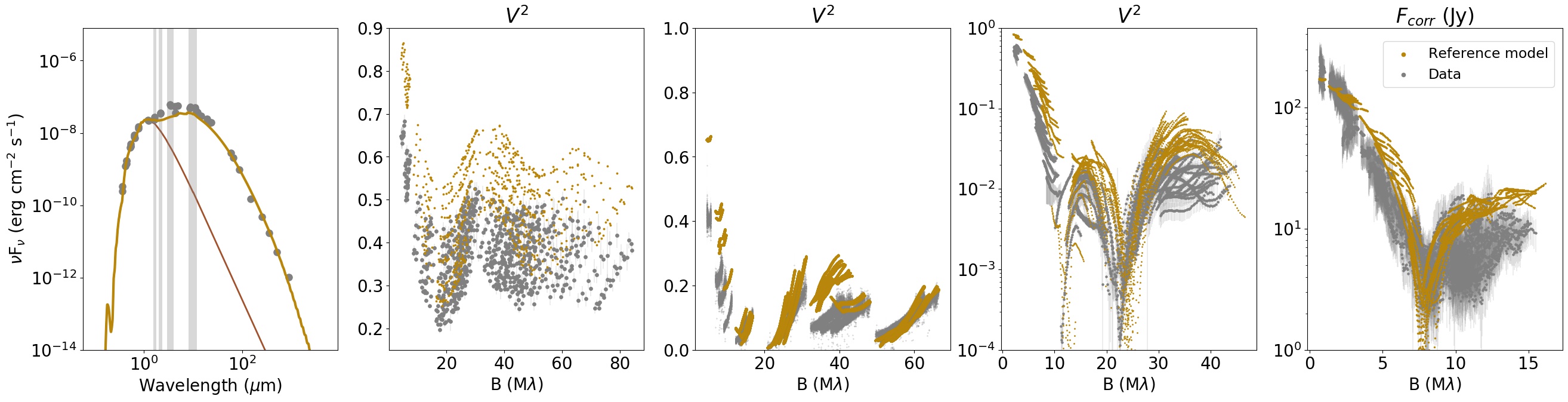}
  \caption{\modiflet{SED and visibility curves of the four interferometric bands of the reference model. 
  From left to right: the reddened SED, the $H$-band, $K$-band, and $L$-band squared visibility curves, and the correlated $N$-band fluxes.}
  The wavelength regimes of the $H$, $K$, $L$, and $N$ bands and the reddened stellar photosphere are depicted in the SED plot by the vertical grey areas and by the brown curve, respectively.
 }
  \label{fig:SEDVIS_reference_model}
\end{figure*}

\subsection{The reference model}
\label{section:referencemodel}
Both the SED and the squared visibility measurements of the PIONIER dataset for IRAS\,08544 were reproduced by \citet{Kluska_2018} using the two-dimensional version of MCMax.
Here, we build upon this model. 
We translated this model to MCMax3D to be able to incorporate azimuthally asymmetric features in the future.
We assumed a distance of $1.22^{+0.01}_{-0.003}$ kpc from the recent \textit{Gaia} data release 3 \citep{Gaia_2016, Gaia_DR3_2022} and rescaled the stellar and disc parameters accordingly. \textit{Gaia} did not flag this object as an astrometric binary and hence this distance is determined using a single star fit to the astrometric data.
The stellar photosphere of the post-AGB star is modelled from the results of the SED fitting by \citet{Kluska_2022}.
Stellar masses from the updated distance estimate are calculated in the same way as the upper limit estimates of \citet{Kluska_2018}.
The mass of the post-AGB star is estimated using the luminosity-core mass relation for post-AGB stars \citep{Vassiliadis_1994}, as it is expected to have lost most of its envelope.
The stellar parameters are reported in Table \ref{Table:Paramsspace}.

As we are interested in the radial and vertical disc structure, we neglect the binary nature of the system in our modelling as the post-AGB primary is much more luminous than the main sequence companion and the binary separation is negligible compared to the size of the disc \citep{Oomen_2018}. We also neglect the effect of the (asymmetric) irradiation of on the disc due to the non-central energy source (see Sect.\,\ref{section:futureprospects} for future implications).
In the post-processing phase, the contribution of the accretion disc around the secondary is taken into account in the photometry by adding a blackbody with a temperature of 4000\,K and a flux contribution of 3.9\% at 1.65\,$\mu$m, which are taken from a fit to the PIONIER data by \citet{Hillen_2016} and \citet{Kluska_2018}.
Likewise, this contribution is added to the synthetic interferometric images by assuming for simplicity that the emission coincides with the position of the primary.


We used the refined model for dust opacity of the DiscAnalysis (DIANA) project \citep{Woitke_2016}. 
Since the circumbinary discs around post-AGB binaries are mostly found to be oxygen-rich \citep[e.g.][]{Gielen_2011}, the carbon fraction is set to zero, leaving the dust to consist of amorphous pyroxene silicates (Mg$_{0.7}$Fe$_{0.3}$SiO$_3$).
The interstellar-medium-like opacities that were assumed in the two-dimensional model were found to best correspond to irregular shaped particles with a porosity, $p$, of 0\% and a distribution of hollow spheres \citep{Min_2005} with a maximum hollow volume ratio, $f_\mathrm{max}$, of 0.7.
We note that these are different from the standard DIANA opacity, as it takes a porosity of 25\% and $f_\mathrm{max}$ = 0.8. 
The opacities are calculated using a \modiflet{standard MRN-like} distribution following \citet{1977ApJ...217..425M}, who showed that the grain size distribution with $q = 3.5$ reproduced the extinction curve in the Milky Way.

We call this adapted version of the previous best-fit model our reference model.
Parameters of this reference model can be found in Table \ref{Table:Paramsspace}.  The performance of this model on the photometry and the IR interferometric datasets are shown in Fig. \ref{fig:SEDVIS_reference_model}.

\begin{figure*}
\centering
   \includegraphics[width=17cm]{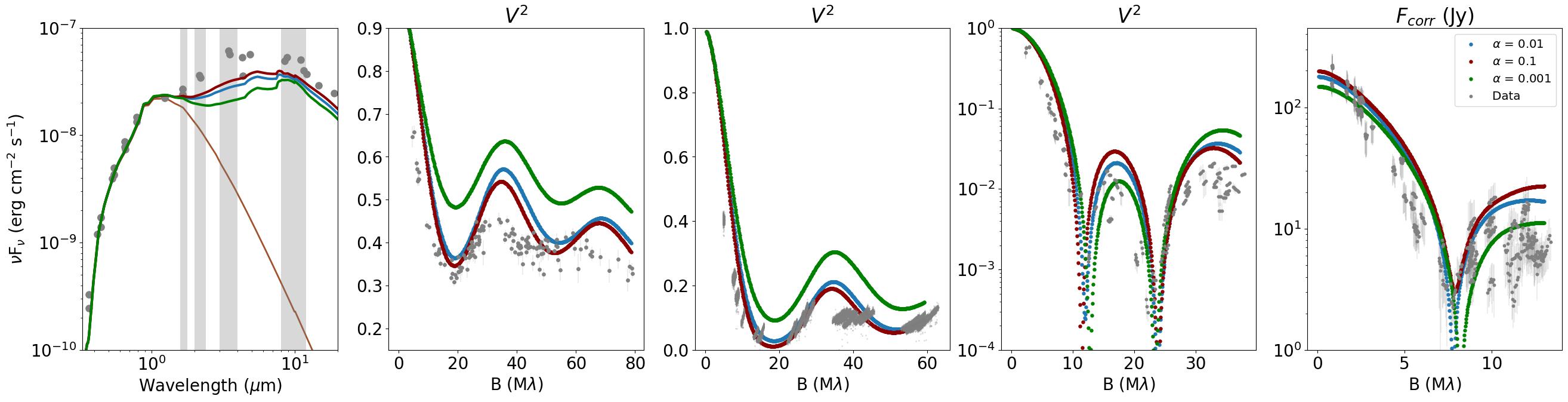}
   \includegraphics[width=17cm]{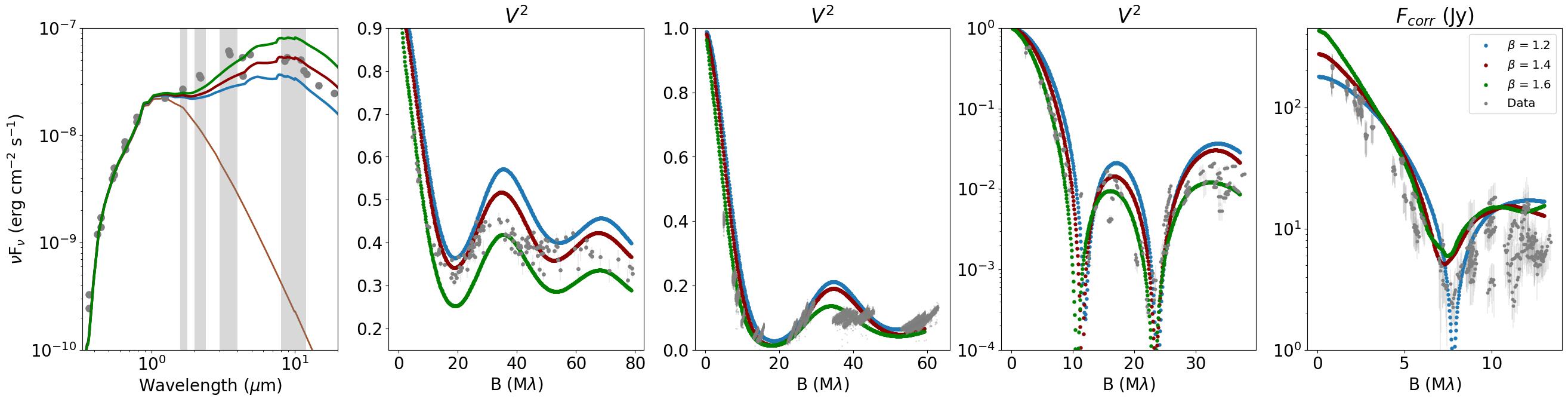}
   \includegraphics[width=17cm]{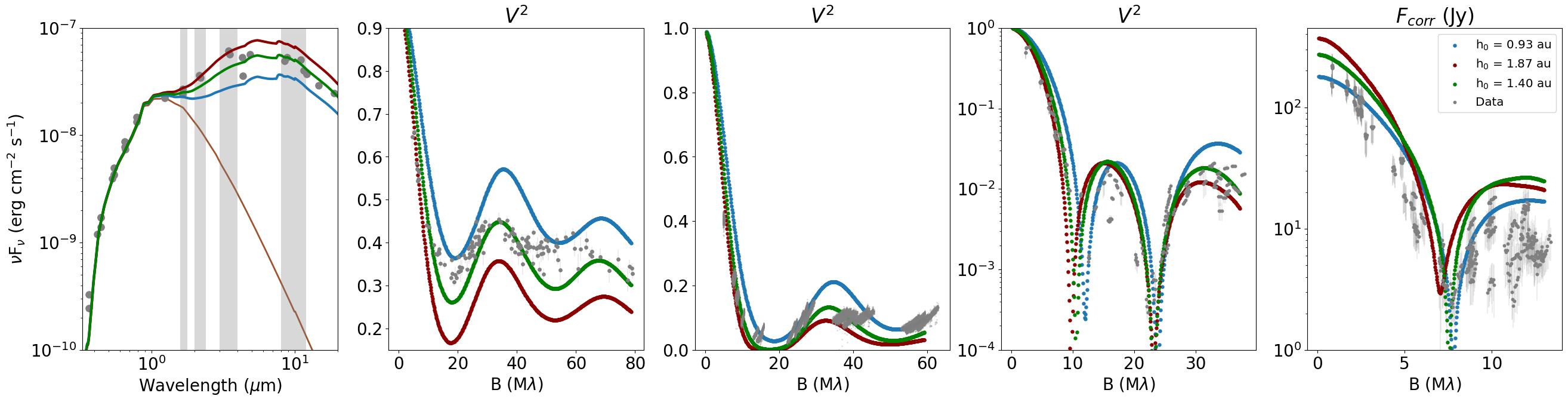}
   \includegraphics[width=17cm]{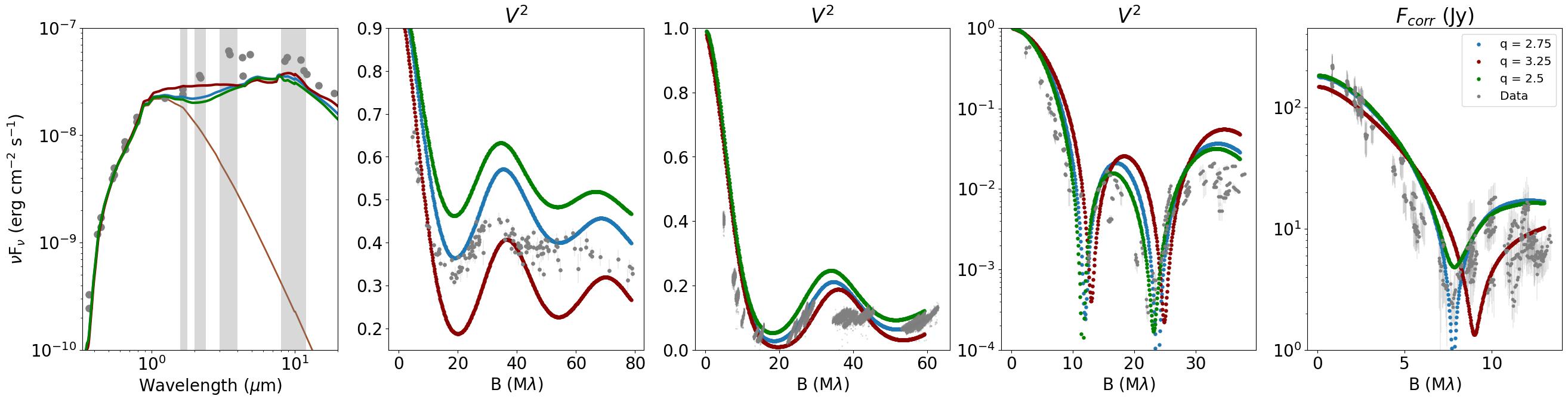}
  \includegraphics[width=17cm]{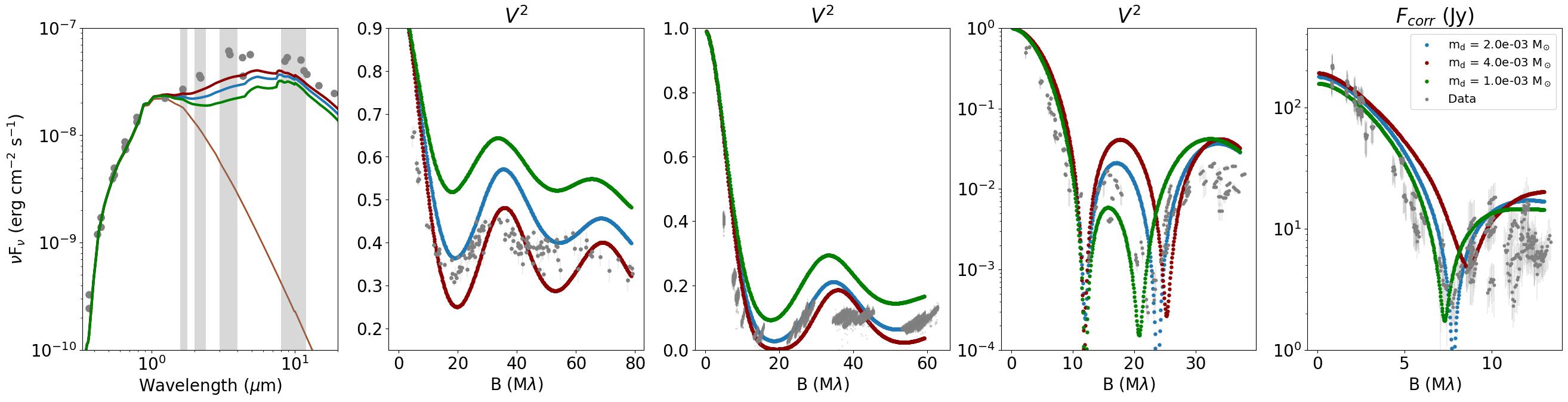}
   
  \caption{Variations in the SEDs and visibility curves for the most impactful parameters.
  From left to right: the reddened SEDs, $H$-band squared visibilities, $K$-band squared visibilities, $L$-band squared visibilities, and correlated $N$-band fluxes.
  The SED plots zoom in to the area of interest (i.e. in the wavelengths for which we have IR interferometric data) to highlight the changes.
  The reddened stellar photosphere is depicted in brown in the SED plots.
  Here, the wavelength regimes of the $H$, $K$, $L$, and $N$ bands are depicted by the vertical grey areas.
  The reference model is always depicted in blue.
  The visibilities are calculated at synthetic baselines from 1-150\,m.
  Interferometric data are only shown around the central wavelengths of the wavelength bands.
  From top to bottom: variations in $\alpha$, $\beta$, $h_0$, $q$, $m_\mathrm{d}$, and $r_\mathrm{mid}$.
  }
  \label{fig:parameters_SEDVIS_mostimpacting}
\end{figure*}
     
\begin{figure*}
\ContinuedFloat 
\centering
     \includegraphics[width=17cm]{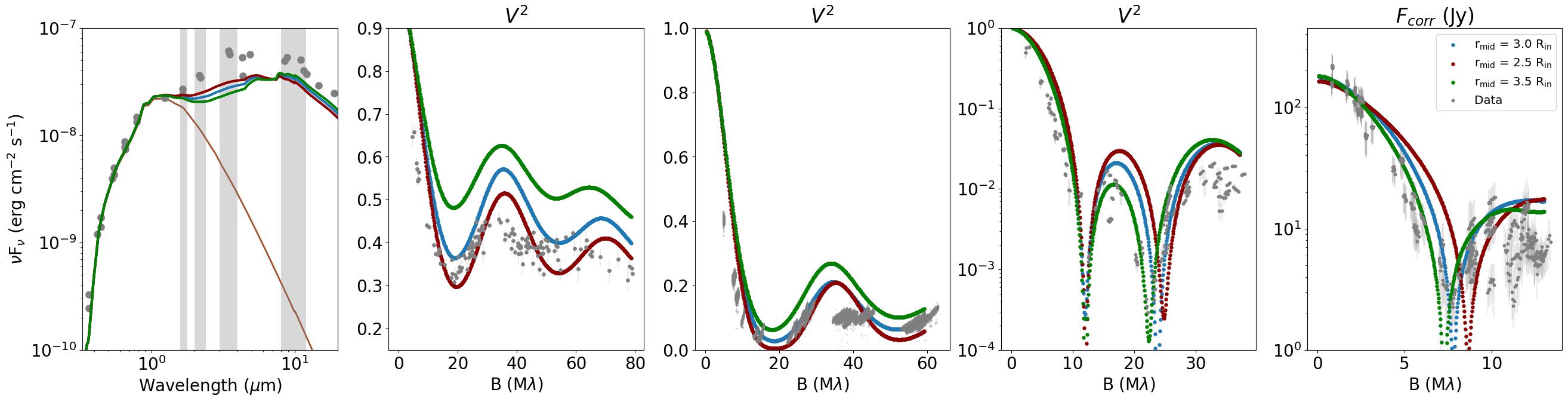}
  \caption{Cont. Variations in $r_\mathrm{mid}$.
  }
\end{figure*}

%

\begin{figure*}
\centering
   \includegraphics[width=17cm]{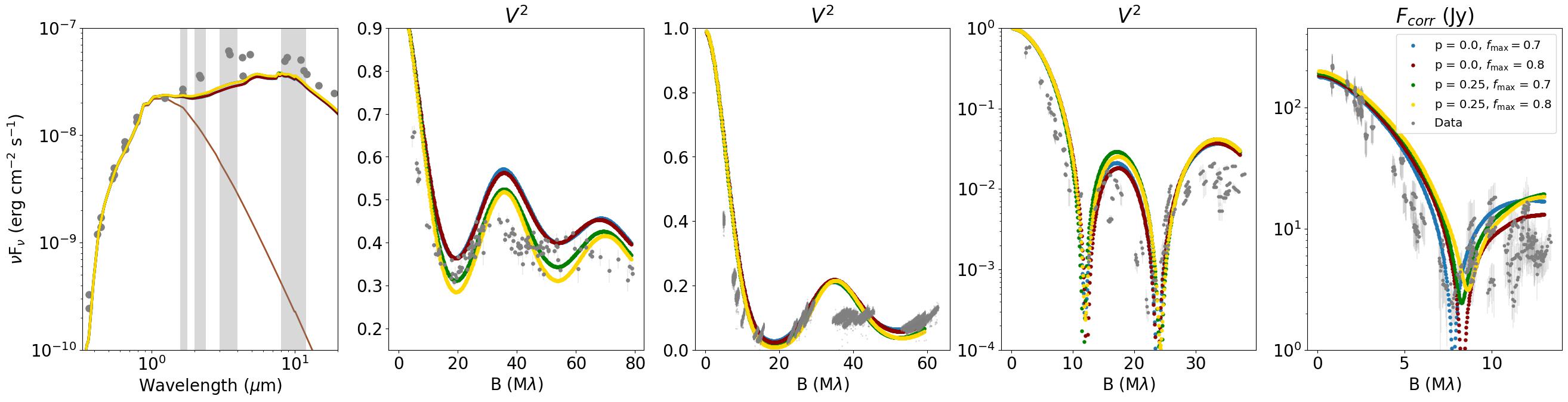}
   \includegraphics[width=17cm]{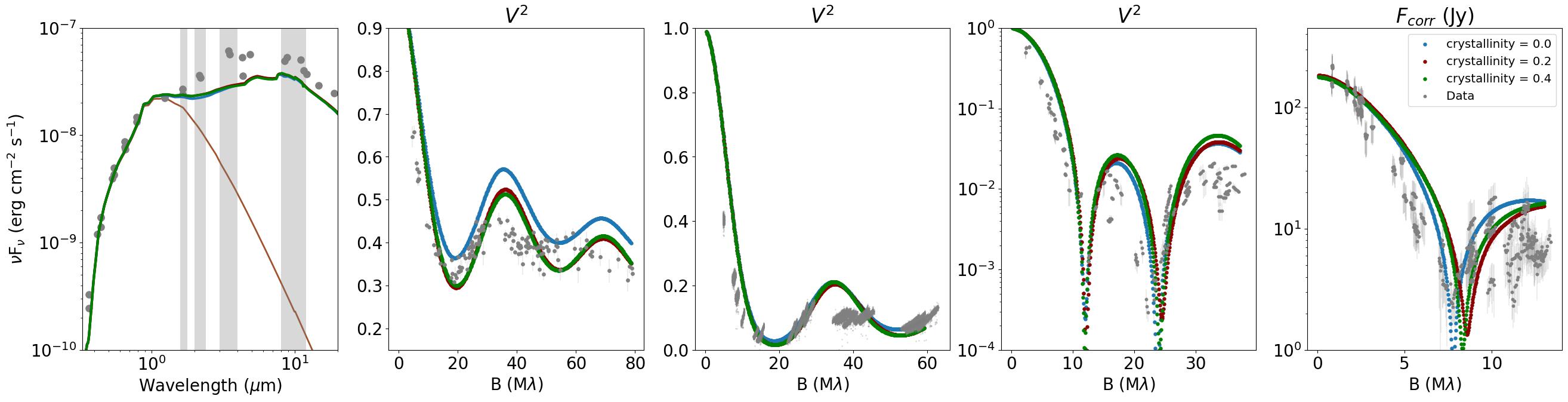}
   \includegraphics[width=17cm]{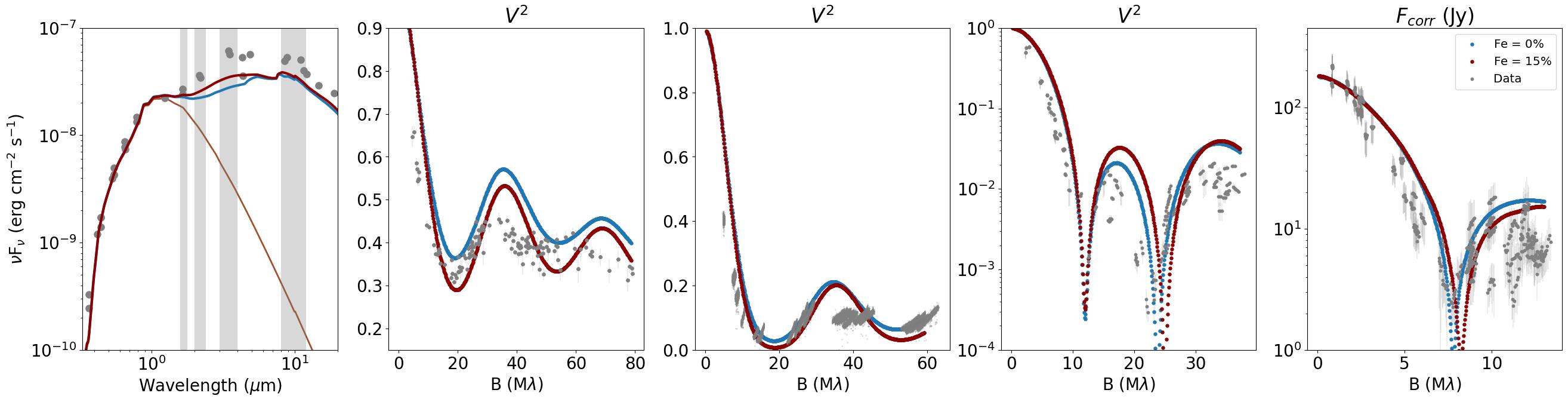}
  \caption{Same as Fig. \ref{fig:parameters_SEDVIS_mostimpacting} but for the dust composition.
  \textit{Top.} Effect of changing the porosity and the maximum hollow volume ratio assuming the DIANA project opacities.
  \textit{Middle.} Effect of adding a crystalline component.
  \textit{Bottom.} Effect of adding a metallic iron component.}
  \label{fig:dustcomposition-visibilities}
\end{figure*}

\begin{table}
\caption{Parameter space for the parametric study. }

\label{Table:Paramsspace}
\begin{tabular}{lccccc}
\hline 
Stellar properties \tablefootmark{a}\\ 
Parameter & Value \\
\hline
$T_\mathrm{eff}$ (K) & 7250\\
$\log g$ (dex) &   1.01 \\
$R_\mathrm{Post-AGB\,star}$ (R$_\odot$) & 65 \\
$L_\mathrm{Post-AGB\,star}$ (L$_\odot$) & 10500 \\
$d$ (kpc) & 1.22 \\

\hline
Parameter &reference model & grid range\\
\hline
$\alpha$& 0.01 &[0.001, 0.01, 0.1] \\
$\beta$ & 1.2 & [1.2, 1.4, 1.6] \\
a$_\mathrm{min} (\mu$m) & 0.1 & [0.01, 0.1, 1.0] \\
gas/dust & 100 & [50, 100, 200] \\
$h_0$ (au) & 0.93 & [0.93, 1.40, 1.87]\\
$m_{d}$ ($\times 10^{-3}$ M$_\odot$) & 2.0 & [1.0, 2.0, 4.0] \\
$p_\mathrm{in}$ & -1.5 & [-1.0, -1.5, -2.0]\\
$q$ & 2.75 & [2.50, 2.75, 3.25] \\
$r_\mathrm{mid}$ (R$_\mathrm{in}$) & 3.0 & [2.5, 3.0, 3.5]\\
$R_\mathrm{in}$ (au) & 7.20 & fixed\\
$R_\mathrm{out}$ (au) & 175 & fixed\\
$a_\mathrm{max}$ (mm) & 1.0 & fixed \\
$i$ (deg) & 19 & fixed\\
PA (deg) & 6 & fixed\\
$p_\mathrm{out}$ & 1.0 & fixed\\

\hline
Dust composition &reference model & grid range\\
\hline
porosity & 0\% & [0\%, 25\%] \\
$f_\mathrm{max}$ & 0.7 & [0.7, 0.8] \\
crystallinity fraction & 0.0 & [0.0, 0.2, 0.4]\\  
metallic iron & 0\%& [0\%, 15\%] \\
\hline
\\

\end{tabular}
\tablefoot{ \tablefootmark{ a} Values for the displayed stellar parameters are taken from \citet{Kluska_2022} and \citet{Kluska_2018} and re-scaled with the distances from \textit{Gaia} DR3 \citet{Gaia_DR3_2022}.
}
\end{table}

 \subsection{The \modiflet{over-resolved} component}
  \label{section:strategy-extended_component}

The model of \citet{Kluska_2018} needed an additional ad hoc \modiflet{over-resolved} component contributing 8.1\% in the $H$ band to fit the SED and reproduce the $H$-band interferometric measurements. 
This component points towards an over-resolved emission that is not reproduced by the reference model.
Indeed, Fig. \ref{fig:SEDVIS_reference_model} shows that by excluding this \modiflet{over-resolved} component, both the photometric and the interferometric data are not well reproduced.
The model visibility curve in the $H$ band lies above the PIONIER data as a result of this missing flux. 
This offset indicates that the stellar flux relative to the total is over-estimated.
The model lacks photometric fluxes in the near-IR and mid-IR, while the millimetre fluxes are fitted well.
For these reasons, we aim at finding models that have more photometric fluxes in all bands and more over-resolved flux in the $H$, $K$, and $L$ bands.

One possibility to explain this over-resolved emission is that it is a scattering component coming from the outer disc and hence originating from a flared disc.
For a more flared disc, we expect the outer regions of the disc to intercept more starlight and thus lead to more scattering.

Besides the over-resolved flux component, the reference model fails to fit the visibilities at short baselines before the first zero, which provides a measure for the wavelength-dependent radial profile.
This leads to a significant underestimation of the size of the emission in these bands.


\subsection{Strategy}
\label{section:strategy-parameter}
In this section we define how we set up the exploration of the parameters while showing the results in Sect.\,\ref{section:results-parameters}.

\subsubsection{Exploration of the parameter space}
\label{section:strategy-exploration}
Since the effect of the different parameters has not been examined extensively for such post-AGB circumstellar discs yet, we start by exploring our parameter space by investigating (1) the disc geometry with the primary aim to study the \modiflet{over-resolved} component and (2) the radial profile of the disc within our interferometric observations. 

We varied various disc parameters based on previous works in circumstellar disc modelling \citep[e.g.][]{Woitke_2016}.
Table \ref{Table:Paramsspace} lists the disc parameters along with the grid range for the parametric study.
In general, we tested values that are both higher and lower than the reference model values except for parameters influencing the amount of the over-resolved emission such as $\beta$ and $h_0$ (Eq. \ref{Eq:scaleheight}) that control the flaring.
The IR interferometric observations are not sensitive to the maximum grain size, and the outer disc radius.
These parameters are therefore fixed to one value throughout this study.
The inner disc radius is well constrained by the PIONIER data \citep{Kluska_2018} and is kept fixed to their best-fit value, scaled to the updated distance from \textit{Gaia} DR3, of 7.2\,au.
The inclination, $i$, and position angle, $PA$, are constrained by the geometric model fitting of the PIONIER data by \citet{Hillen_2016}. 
We kept these parameters fixed to $i = 19^\circ$ and $PA = 6^\circ$ (measured north to east), respectively.

The mid-IR interferometric observations are also sensitive to the radial opacity profile. 
To investigate the effects of the opacity on the observables, we varied three dust properties.
First, we varied the porosity and the $f_\mathrm{max}$ of the amorphous silicate dust mixture, as determined by the DIANA project, and adopted the values of the DIANA standard.
Second, we added a crystalline component by taking a mixture of the amorphous silicates of the DIANA project and crystalline forsterite.
Forsterite was chosen since it was found that it is the dominant source of crystalline dust in discs around post-AGB binaries \citep{Gielen_2008}. 
The opacities are computed from the optical constants of \citet{Servoin_1973} assuming the grain properties are in agreement with our standard dust mixture (i.e. with a porosity of 0\% and $f_\mathrm{max} = 0.7$).
For the crystalline component, we tested crystallinity fractions of 0.2 and 0.4 by volume.
Third, we mixed the amorphous silicates with metallic iron.
The formation of metallic iron in the disc environment has been debated in, for example, \citet{Hillen_2014}.
We tested a mixture of 85\% amorphous silicate with 15\% metallic iron by mass. 
For metallic Fe, optical constants of \citet{Palik_1991} were used to compute the opacities. 
In all cases the dust composition is assumed to be homogeneous throughout the disc. 

To investigate the impact of the \modiflet{model} parameters on the radial morphology of the emission, we used the half-light radii (hlr) of the disc in different spectral bands.
This quantity specifies the radius at which half of the flux emitted at a given wavelength is captured.
These hlr are constrained by geometric models in \citet{Corporaal_2021}.

\begin{figure*}
\centering
 \begin{minipage}{0.33\textwidth}
  \includegraphics[width=\textwidth,width=1.0
  \textwidth]{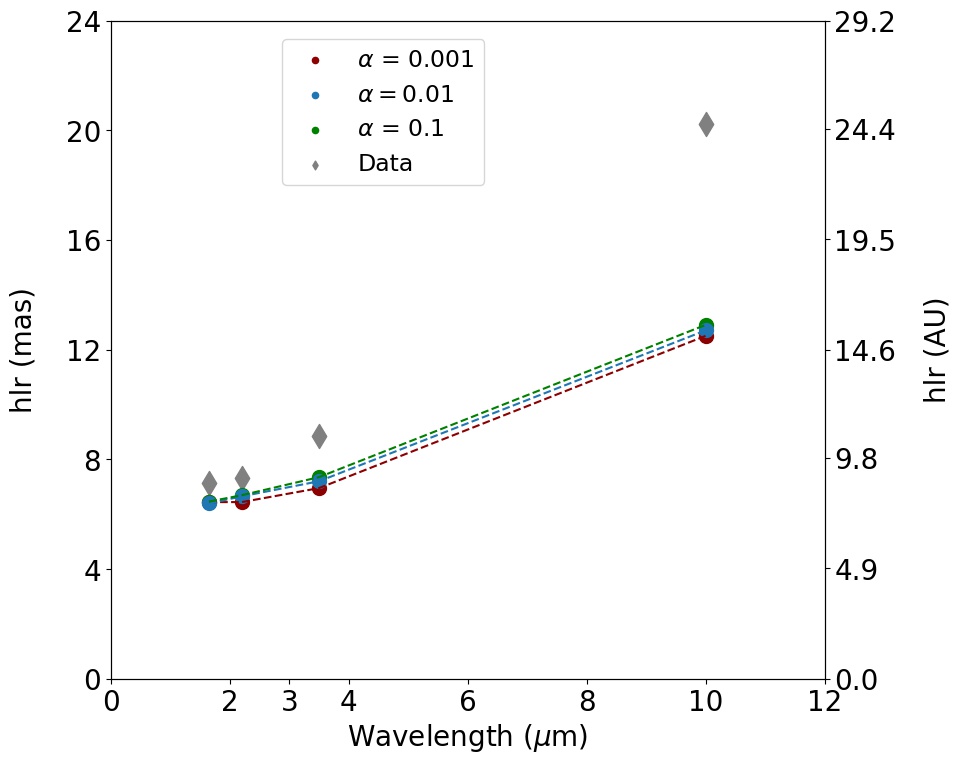}
  \end{minipage}
  \begin{minipage}{0.33\textwidth}
  \includegraphics[width=\textwidth,width=1.0
  \textwidth]{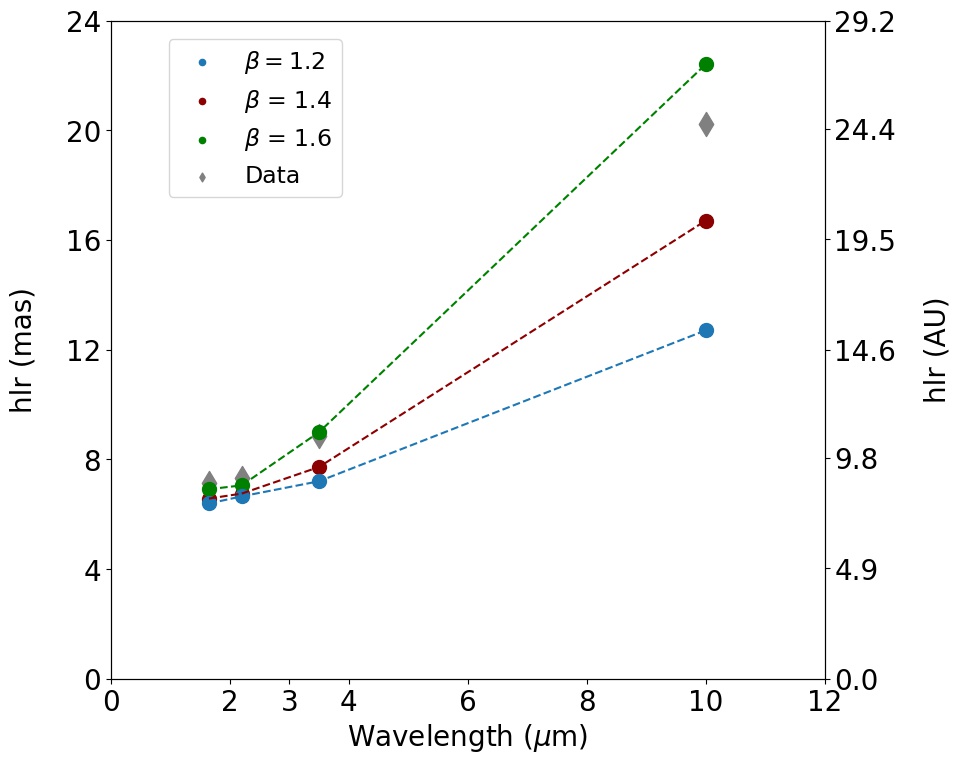}
  \end{minipage}
  \begin{minipage}{0.33\textwidth}
  \includegraphics[width=\textwidth,width=1.0
  \textwidth]{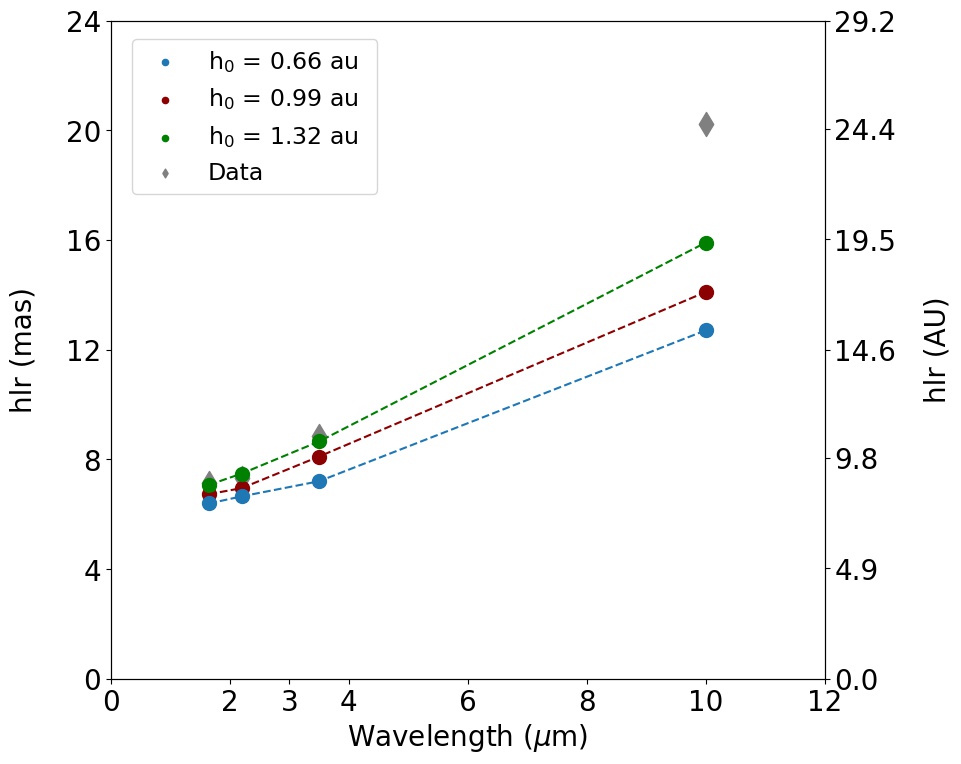}
  \end{minipage}
  
   \begin{minipage}{0.33\textwidth}
  \includegraphics[width=\textwidth,width=1.0
  \textwidth]{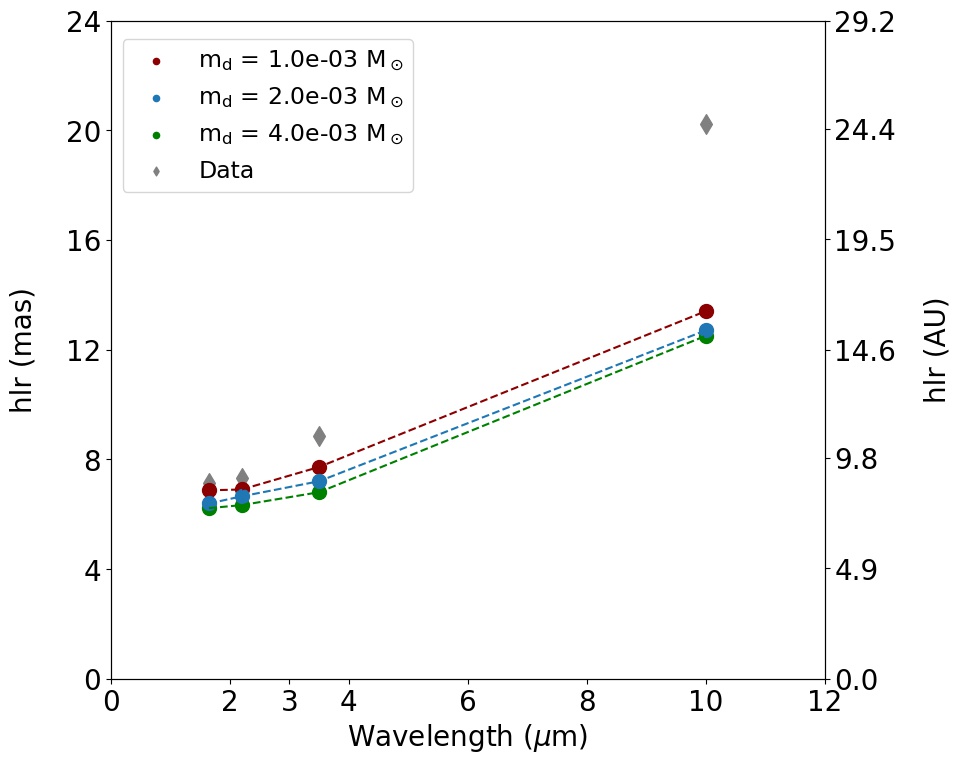}
  \end{minipage}
\begin{minipage}{0.33\textwidth}
  \includegraphics[width=\textwidth,width=1.0
  \textwidth]{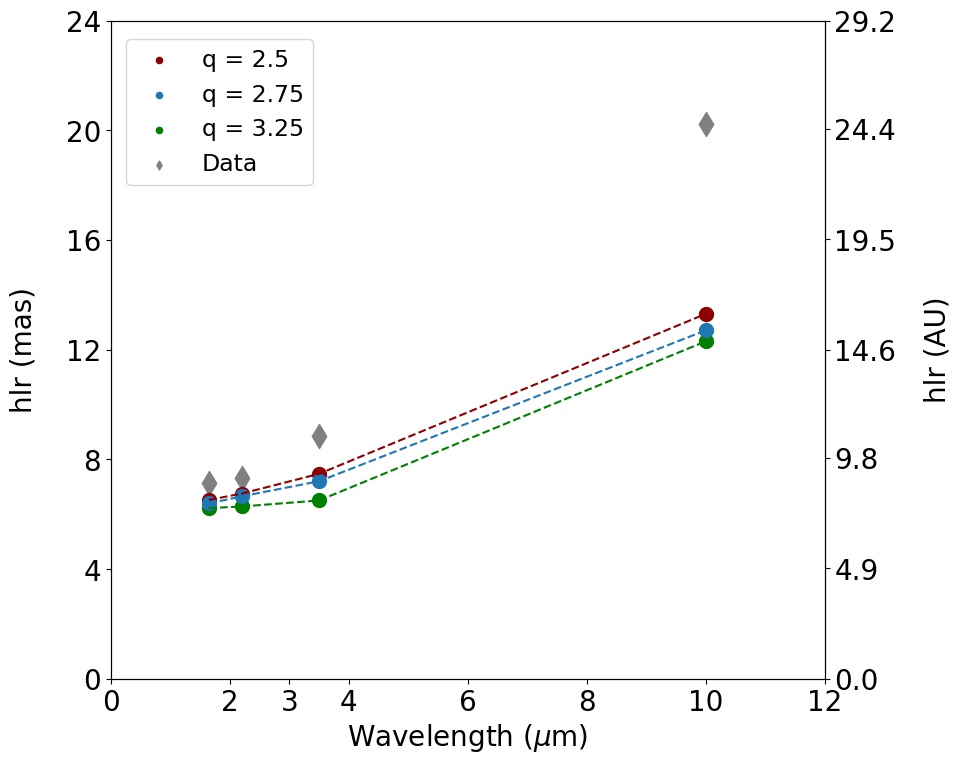}
  \end{minipage}
  \begin{minipage}{0.33\textwidth}
  \includegraphics[width=\textwidth,width=1.0
  \textwidth]{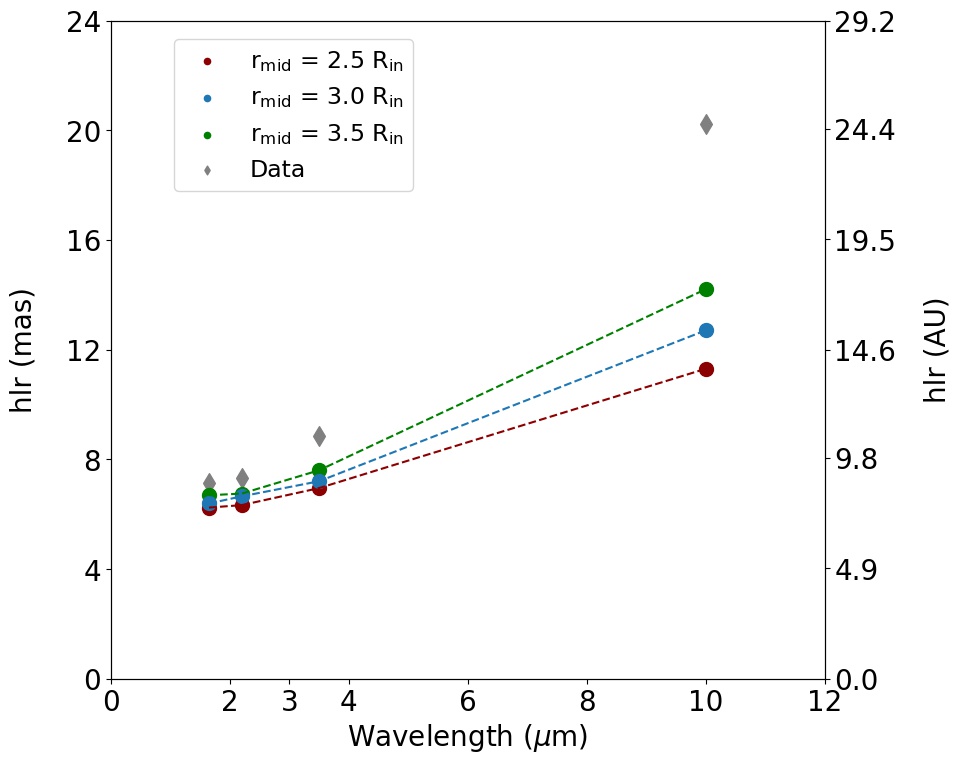}
  \end{minipage}
     \begin{minipage}{0.33\textwidth}
  \includegraphics[width=\textwidth,width=1.0
  \textwidth]{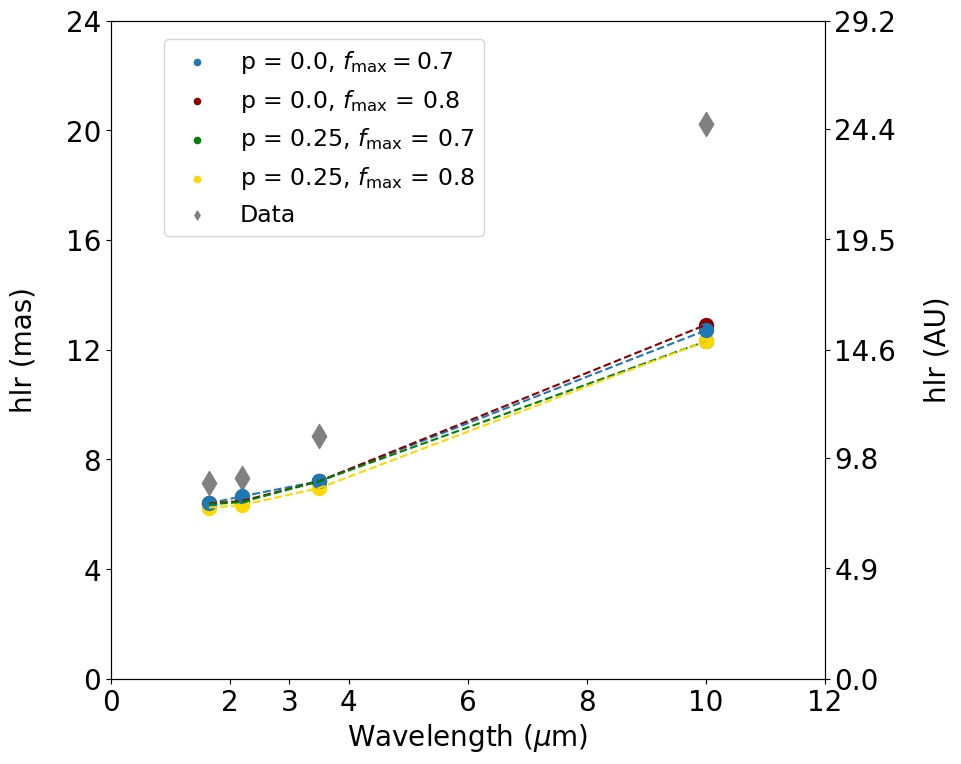}
  \end{minipage}
\begin{minipage}{0.33\textwidth}
  \includegraphics[width=\textwidth,width=1.0
  \textwidth]{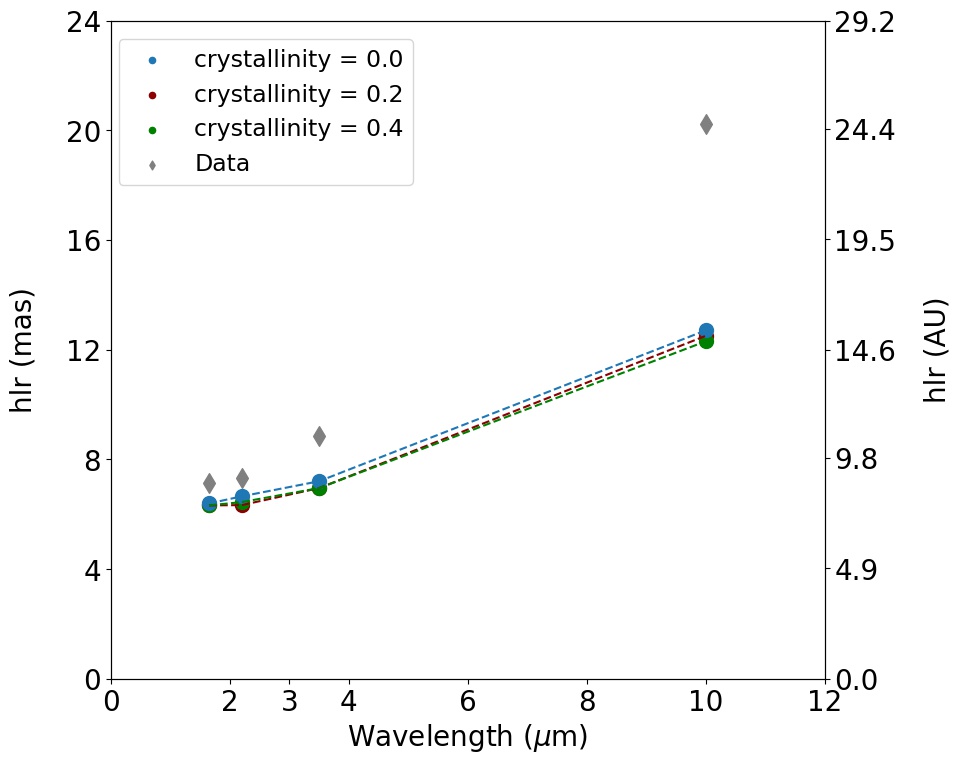}
  \end{minipage}
  \begin{minipage}{0.33\textwidth}
  \includegraphics[width=\textwidth,width=1.0
  \textwidth]{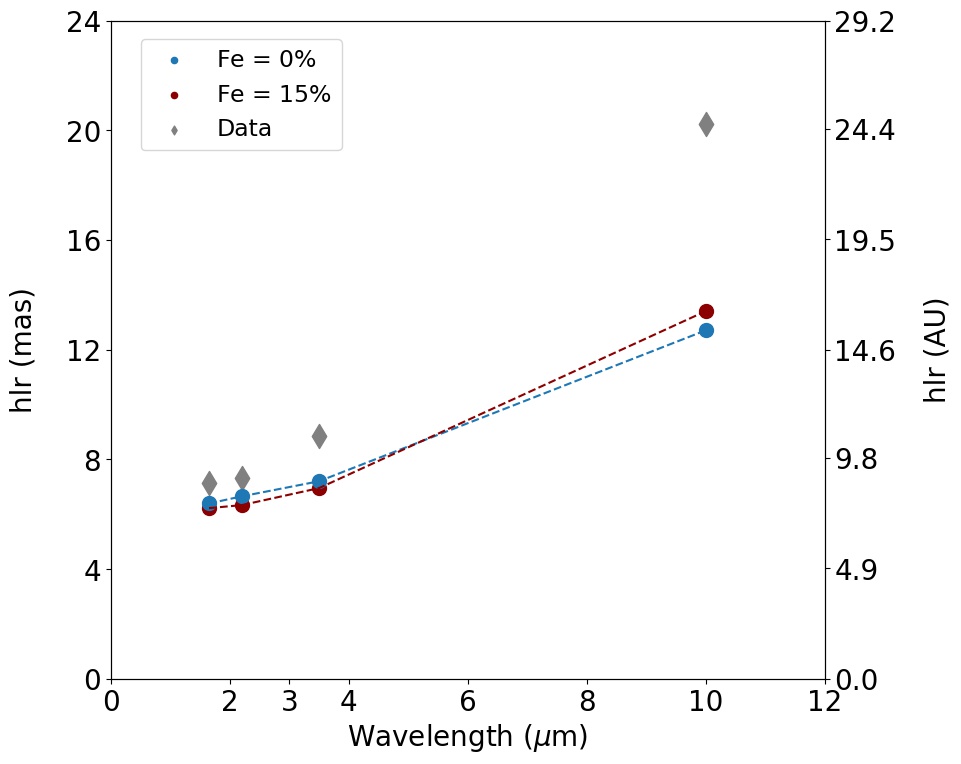}
  \end{minipage}

  \caption{Half-light-radius variations in the radial emission as a function of wavelength for changes in the parameters that have the most promising impact on the photometric and interferometric observables (\textit{top} and \textit{middle}) and for changes in the dust composition (\textit{bottom}).
  The hlr were calculated at the central wavelengths of the $H$, $K$, $L$, and $N$ bands.
  The geometric model data points are taken from \citet{Corporaal_2021}.
  \textit{Top.} Effect of changing $\alpha$ (left), $\beta$ (middle), and $h_0$ (right).
 \textit{Middle.} Effect of changing $m_\mathrm{d}$ (left), q (middle), and $r_\mathrm{mid}$ (right).
  \textit{Bottom.} 
  Variations in the DIANA parameters (left), the crystallinity fraction (middle), and the metallic iron content (right).
 }
  \label{fig:hlr_5impacting_params}
\end{figure*}

\begin{figure*}
\centering
  \includegraphics[width=17cm]{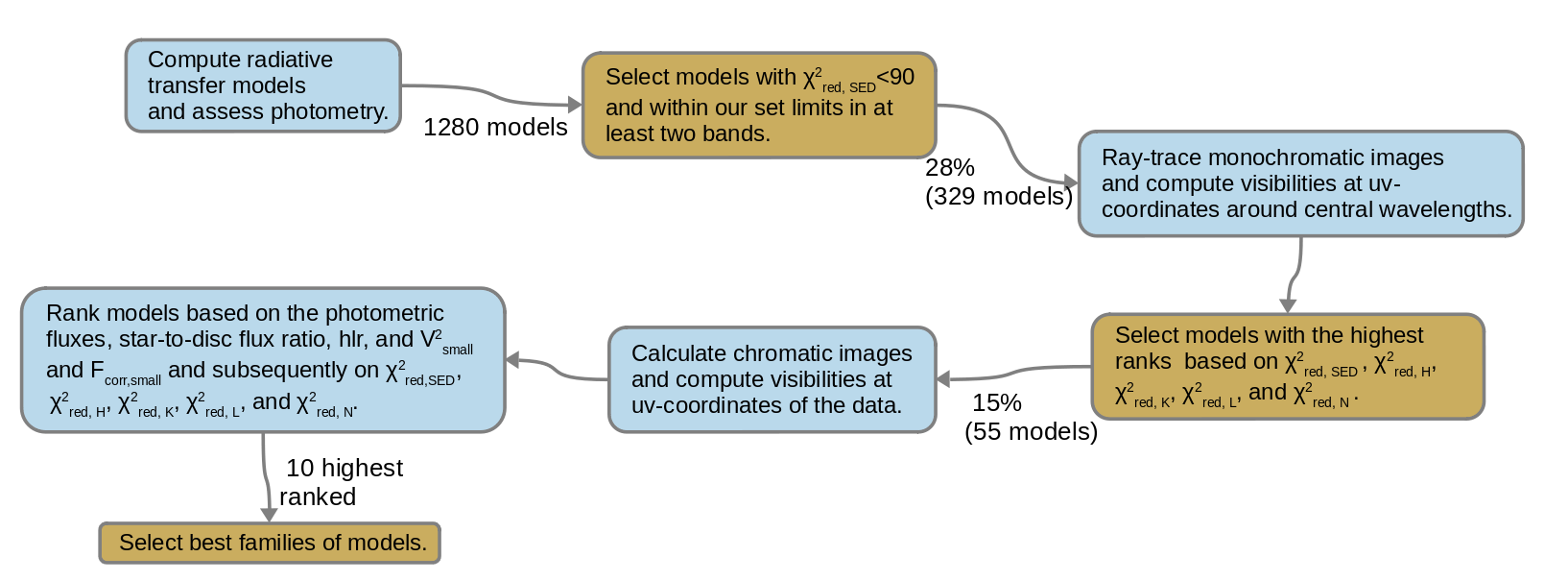}
  \caption{Flowchart representing the different steps taken to find the models that best represent the photometric and interferometric data of IRAS\,08544.
  The blue boxes describe different calculation steps, and the orange boxes indicate the three selection steps.
 }
  \label{fig:flowchart}
\end{figure*}
\subsubsection{Ray-traced spectra and images}
We fitted the interstellar reddening to the photosphere of the post-AGB star for the reference model, by using the Levenberg-Marquardt fitting routine from the Python package lmfit (for a more thorough explanation of the applied reddening, see Appendix\,\ref{appendix:reddening}).
The SED of each model \modiflet{was} then reddened with this value. 

Ray-tracing images is computationally expensive. 
For this reason, we computed images at a single continuum wavelength coinciding with the centres of the wavelength bands of interest.
We compared the image to one given spectral channel of the near-IR interferometric data to avoid any intra-band chromatic effects.

To visualise the effect of varying the parameters on the interferometric visibilities, we plot the squared visibilities for synthetic baselines from 1-150 metres, to match the baseline range of the VLTI. 
Since IRAS\,08544 is almost pole-on,
the direction to which the \textit{uv} coordinates are calculated, does not change the visibility signal significantly.
The synthetic \textit{uv}-coordinate space was calculated along the semi-major axis in the image plane. 
Output images were converted to complex visibilities via a Fourier transform at this synthetic \textit{uv}-coordinate space.
These complex visibilities were subsequently converted to squared visibilities for the $H$, $K$, and $L$ bands and to correlated fluxes in the $N$ band, in agreement with the observables of the data.

\begin{table}[t]
\caption{Explored set of parameters for the application to IRAS\,08544 and the parameters of the family of best-fit models (see also Table \ref{Table:models_10best}).}
\label{Table:paramspace_2}
\centering
\begin{threeparttable}
\begin{tabular}{lccc}
\hline 
Parameter & Grid range & Values best models \\

\hline
$\alpha$& [0.01, 0.1] & [0.1, 0.01] \\
$\beta$ &  [1.2, 1.3, 1.4] & [1.2, 1.3, 1.4]\\
$h_0$ (au) & [0.93, 1.40, 1.87] & [1.40, 1.87]\\
$m_{d}$ ($\times 10^{-3}$ M$_\odot$) & [1.0, 1.5, 2.0, 3.0] & [1.0, 1.5, 2.0]\\
$q$ & [2.75, 3.0, 3.25] & [2.75, 3.0]\\
$r_\mathrm{mid}$ & [2.5, 3.0, 3.5] & [2.5, 3.0, 3.5] \\ 
metallic iron & [0\%, 15\%] & [0\%, 15\%] \\

\hline

\end{tabular}
\end{threeparttable}
\end{table}

\subsection{Results of parametric study}
\label{section:results-parameters}
The impact on the SED and on the visibilities for the most impactful structural parameters of our disc model is illustrated in Fig. \ref{fig:parameters_SEDVIS_mostimpacting}. In Fig.~\ref{fig:dustcomposition-visibilities} we show the same but for variations in the dust composition. 
The impact of these parameters on the hlr are shown in Fig. \ref{fig:hlr_5impacting_params}.
The impact on the photometry and interferometry and on the hlr for a change in the other parameters are shown in Figs. \ref{fig:parameters_SEDandVIS_rest} and \ref{fig:hlr_rest_params}, respectively.



\subsubsection{The impact of the turbulence parameter}
The turbulence parameter controls the strength of the dust settling and, therefore, the scale height of the larger grains. 
Stronger dust settling leads to the removal of large grains from the surface layers.
A decrease in $\alpha$ decreases the photometric fluxes, increases the star-to-disc flux ratio significantly and increases the over-resolved flux component.
An increase in $\alpha$ shows, however, only slight variations with respect to the reference model, such as a slight increase in the photometric fluxes, suggesting that the number density of small grains where IR interferometric observations are sensitive to are not changing from $\alpha = 0.01$ to $\alpha = 0.1$.
Neither an increase nor a decrease in $\alpha$ has notable effects on the hlr.

\subsubsection{The impact of the scale height}
The disc flaring as controlled by $\beta$ and $h_0$ strongly affects all observables that we investigated.
Larger $\beta$ and $h_0$ 
significantly increase the near-IR and mid-IR photometric fluxes, as well as the over-resolved emission and the hlr.
For \modif{more flared discs or more vertically extended discs}, the star-to-disc flux ratio is decreased with respect to the reference model, as \modif{these parameters increase} the captured IR flux of the disc due to scattering and absorption. Both parameters impact similarly with two important exceptions: first, they have different effects on the amplitude of the visibility bump in the $N$ band.
Indeed, for increased values of $h_0$, the amplitude increases while this amplitude remains unchanged for varying $\beta$.
Second, increased values of $\beta$ affect the radial extent of the $N$ band more than they affect the radial extents of the $H$, $K$, and $L$ bands.
The model with $\beta=1.6$ is also the only model in which the $N$-band extent deduced from the data is reached, while it also performs well for the near-IR hlr.



\subsubsection{The impact of grain-size distribution}
Variations in the value of $q$ changes the dust grain size distribution between the number of small and large grains, with sizes below and above 1$\mu$m, respectively. 
Varying $q$ leads to changes in the mean dust grain size.
These alter the SED and visibilities significantly.
If the number of small grains is increased with respect to the large grains, or equivalently, if higher values of $q$ are taken, the radial extents in all bands decrease, the SED fluxes increase in the $H$, $K$, and $L$ bands, and the over-resolved emission increase in all bands.
Moreover, the star-to-disc flux ratio is then significantly decreased. 
The small grains contribute considerably to the total flux of the disc such that the disc emission becomes more dominant relative to the star.
A decrease in $q$ slightly decreases the fluxes in the $H$, $K$, and $L$ bands, increases the star-to-disc flux ratio, increases the hlr, and increases the over-resolved flux component. 
\modiflet{The increase or decrease} in the near-IR photometric flux is stronger than in the mid-IR, which is different than what we find for other parameters.
The grain-size distribution thus affects both the radial intensity profile and the emission morphology.

\subsubsection{The impact of the dust mass}
The dust mass impacts the whole SED, the visibilities, as well as the hlr.
Increasing (respectively, decreasing) the dust mass decreases (increases) the stellar contribution significantly and thus shifts the visibility curve in the near-IR as the dust mass in the inner regions is doubled (halved) as well. 
For this reason, higher dust masses decrease the IR excess of the disc and slightly increase the $N$-band over-resolved flux and \modiflet{vice versa} for smaller dust masses.
In the $H$, $K$, and $L$ bands, the over-resolved flux is not altered by varying the dust mass.


\subsubsection{The impact of the turn-over radius}
Variations in the turn-over radius, $r_\mathrm{mid}$, were compared by putting it closer to or farther from the inner rim while keeping the total dust mass constant. 
To ensure continuity in the surface density profile, the ratio of dust mass in the inner zone and the outer zone is altered.
For larger $r_\mathrm{mid}$, the surface density in the inner regions is higher, 
while for smaller $r_\mathrm{mid}$ the turn-over is at smaller radii thus with a lower surface density.
Variations in $r_\mathrm{mid}$ significantly affects all observables.
Putting $r_\mathrm{mid}$ closer to (respectively, farther from) the inner rim decreases (increases) the SED fluxes in the $H$, $K$, and $L$ bands, decreases (increases) the hlr in all bands, and increases (decreases) the star-to-disc flux ratio.

\subsubsection{The impact of the dust composition}
\label{section:parameterstudy_dustcomposition}

The dust composition determines the radial opacity profile of the disc.
Variations in the DIANA opacities show a decrease in the star-to-disc flux ratio and a slight decrease in the hlr for models with $p=0.25$.
Changes in $f_\mathrm{max}$ do not show notable effects on the interferometry.
The crystallinity factor decreases the star-to-disc flux ratio and shows the emission in all bands is similarly radially extended as the reference model.
The addition of metallic iron slightly increases the star-to-disc flux ratio, slightly decreases the radial extents in the near-IR and increases the radial extents in the mid-IR.
Different from the DIANA opacity changes and the addition of the crystallinity factor, the addition of metallic iron increases the photometric flux in the $K$ and $L$ bands significantly, while the $H$- and $N$-band fluxes remain not significantly affected.
For all models with changing dust composition, the over-resolved flux component remains unchanged with respect to the reference model and the hlr values are also barely affected.


\begin{table}
\caption{Parameters of the family of best models and the score, sc, that these models got based on the number of characteristics that the model can reproduce.}
\label{Table:models_10best}
\begin{threeparttable}
\begin{tabular}{p{0.45cm}cccccccp{0.2cm}}
 \\
\hline
 Model& $\alpha$ &$\beta$ & $h_0$  & $m_\mathrm{d}$ &  $q$ &$r_\mathrm{mid}$&Fe & sc
\\
& &&&($\times 10^{-3}$)  \\
&&&(au)& ($\mathrm{M}_\odot$) & &($R_\mathrm{in}$) & (\%)\\
\hline
1 & 0.1 & 1.3 & 1.40 & 1.0 & 2.75 & 2.5 & 0 & 12\\
2 &0.01 &  1.4 &1.40 & 1.0 &2.75 & 2.5 &0  & 11\\
3 & 0.01 & 1.2 & 1.87 & 1.0 &2.75 & 2.5 & 0 &11\\
4 & 0.01 & 1.3 & 1.40 & 1.5 & 2.75 & 3.0 & 0 &11\\
5 & 0.1 & 1.2 & 1.40 & 1.5 & 2.75 & 3.5 & 15 & 11\\
6 & 0.1 & 1.2 & 1.40 & 1.0 & 3.0 & 3.5 & 0 & 10\\
7 & 0.1 & 1.2 & 1.40 & 2.0 & 2.75 & 3.5 & 0 &10\\
8 & 0.1 & 1.2 & 1.40 & 1.0 & 3.0 & 3.0 & 0 &10\\
9 & 0.1 & 1.2 & 1.40 & 1.5 & 2.75 & 3.0 & 0 & 10\\
10 & 0.01 & 1.2 & 1.40 & 1.0 & 3.0 & 3.5 & 15 & 10\\
\hline
\\
\end{tabular}
\end{threeparttable}
\end{table}


\subsection{Selection of the most impactful parameters}
\label{section:selection-parameters}

Here, we outline our selection of the parameters that have most impact on the observables with a specific focus on the over-resolved flux component.
We find the following six structural parameters that have the most impact on the SED and the interferometry in terms of the photometric flux, the over-resolved flux, and the radial extent: $\alpha$, $\beta$, $h_\mathrm{0}$, $m_d$, $q$, and $r_\mathrm{mid}$.
For $\alpha$, $\beta$, $h_\mathrm{0}$, and $q$ we find that increased values of these parameters with respect to the reference model show results \modiflet{that improve} the fit to the SED in the near-IR and mid-IR, the over-resolved flux component, and the radial profile of the disc.
Variations in the other parameters with significant impact, $m_d$ and $r_\mathrm{mid}$, need to be explored in both ways with respect to the reference model.
Therefore, we selected values for these six parameters that are either the values of the reference model or in the direction that improves the fit to the observables (i.e. the SED or the interferometric measurements).

\begin{figure*}
\centering
  \includegraphics[width=17cm]{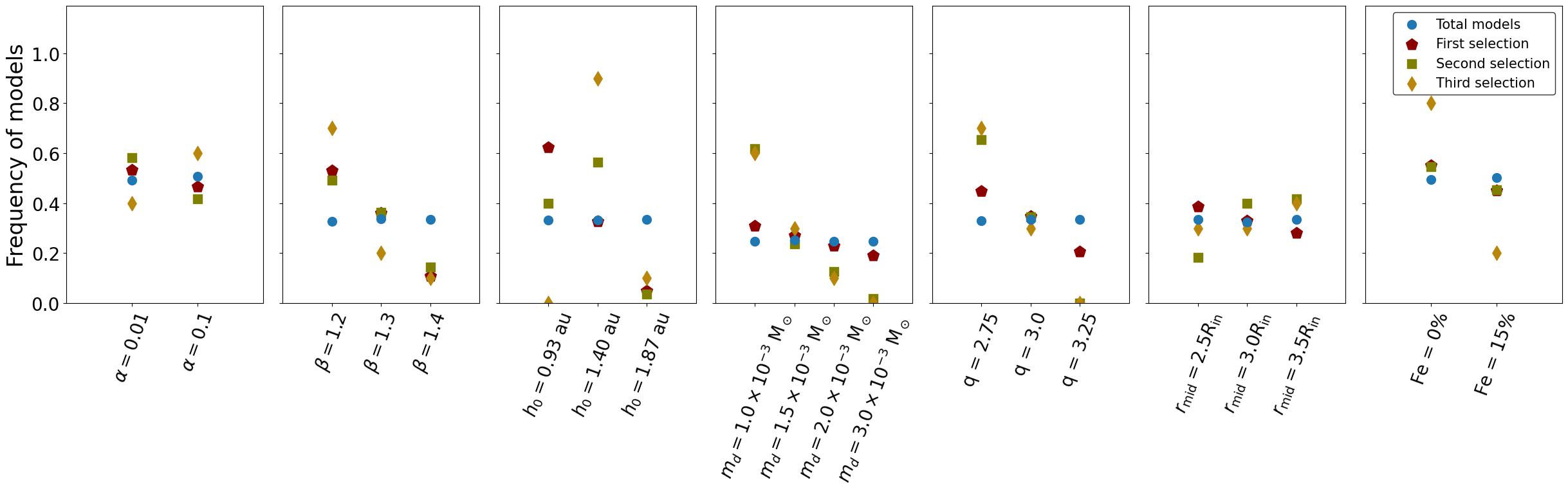}
  \caption{Distribution of the parameter space resulting from each of the three selection steps.
  The parameter space is defined by (from left to right) $\alpha$, $\beta$, $h_0$, $m_d$, $q$, $r_\mathrm{mid}$, and Fe. 
  The distribution for the total number of models (1380 models) is indicated by the circles, which are equally distributed over each of the parameters.
  The final distribution of parameters of our family of best-fit models is indicated by the diamonds.
 }
  \label{fig:selection}
\end{figure*}
Changes in the dust compositions have smaller impacts on the explored observables as compared to the structural parameters.
Effects of the opacity are therefore not considered to play a significant role in explaining the over-resolved component.
Moreover, it remains difficult to break the degeneracies between the effect of the different dust compositions.
However, the addition of metallic iron as an additional opacity source significantly increases the $K$- and $L$-band photometric fluxes, while not changing the morphological characteristics of the emission.
For this reason, we included the metallic iron as a parameter in our grid.
We tested both models assuming a dust composition fixed to the composition of our reference model and models with a dust composition that is a mixture of silicates (85\%) and metallic iron with content (15\%) and kept this fixed throughout the disc.
In what follows, we take these seven parameters for a more in-depth study while we keep all other parameters fixed to the values of the reference model.

\section{Application to IRAS\,08544}
\label{section:application}
In this section we perform a more in-depth study of the characteristics of the circumbinary disc around IRAS\,08544.
We outline our strategy to find our family of best-fit models in Sect.\,\ref{section:strategy-application} and present the results in Sect.\,\ref{section:results-application}. We performed a systematic study within the parameter space derived from the results in \modiflet{the} previous section.
Table \ref{Table:paramspace_2} displays the range of the varied parameters. 

\subsection{Strategy}
\label{section:strategy-application}

To limit computational time, our strategy consists of three selection steps.
A graphical representation of our strategy is depicted in Fig. \ref{fig:flowchart}.
We first ran the radiative transfer models, computed the photometry, and subsequently assessed the performance of the SEDs with respect to the data before creating the images and comparing the model to the interferometric measurements.

From the full investigation of the parameter space, we \modiflet{got} 1280 unique models.
Photometric data points in the $H$, $K$, and $L$ bands \modiflet{were} taken from \citet{deRuyter_2006}.
To avoid discrepancy between the model and the data due to the 11.3\,$\mu$m crystalline forsterite feature, interpolation \modiflet{was} performed between all photometric flux points in the $N$ band (from 8-13\,$\mu$m) to find the photometric flux at 10\,$\mu$m.
This \modiflet{was} then compared to the photometric flux at 10\,$\mu$m of the models. 
The amorphous pyroxene silicate feature at 9.8\,$\mu$m is broad and extends throughout the $N$ band such that we cannot bypass this feature.
We selected confidence intervals in which we considered the model fluxes to be good enough within our framework. 
These are within 20\% of the photometric data in the $H$, $K,$ and $L$ bands and 25\% of the interpolated data point at 10\,$\mu$m in the $N$ band.
These levels take into account the uncertainties in the photometric data points.

We fitted the interstellar reddening for each model spectrum individually (see also Appendix\,\ref{appendix:reddening}).
The SED of each model is then reddened accordingly.
The reduced $\chi^2$ of this reddened SED, $\chi^2_\mathrm{red, SED}$, was calculated by taking into account only photometric data points up to 20\,$\mu$m, since our interferometric observables are sensitive to neither the geometry of the outer disc nor larger grain sizes. 

As a first selection step, we selected models based on the following two criteria:
\modiflet{(1)} models that have $\chi^2_\mathrm{red, SED}$ that are smaller than the $\chi^2_\mathrm{red, SED}$ of the reference model (which has $\chi^2_\mathrm{red, SED}$ = 90) and \modiflet{(2)} models that comply with the observed photometry within our set limits in at least two out of the four bands.
For models that satisfy these criteria, we computed the monochromatic images at the central wavelengths of each band.
To compute the synthetic visibilities, we used the \textit{uv} coordinates of the data and performed a Fourier transform at these \textit{uv} coordinates.

\begin{table*}

\caption{SED characteristics of the family of best-fit models.}
\label{Table:models_SED_10best}
\centering
\begin{threeparttable}
\begin{tabular}{lccccccccccccccccccccccc}
 \\
\hline
 Model& $F_\mathrm{H}$ &  $F_\mathrm{K}$ &  $F_\mathrm{L}$ & $F_\mathrm{N}$ & $\chi^2_\mathrm{red, SED}$  \\
& ($\times 10^{-8}$) & ($\times 10^{-8}$)  & ($\times 10^{-8}$) & ($\times 10^{-8}$)\\
&(erg/s/cm$^2$) & (erg/s/cm$^2$) &  (erg/s/cm$^2$)   &  (erg/s/cm$^2$)\\
\hline
data &2.60 & 3.46 & 5.48 & 5.00 & - \\
1 & $\bf{2.50}$ &  $\bf{2.96}$ &  $\bf{4.59}$ &  $\bf{5.49}$ &   45 \\

2 & $\bf{2.52}$ &  $\bf{2.86}$ &         4.32 &  $\bf{5.77}$ &   55  \\
3 & $\bf{2.63}$ &  $\bf{3.16}$ &  $\bf{4.79}$ &  $\bf{5.27}$ &   38 \\

4 &  $\bf{2.48}$ &         2.76 &         4.16 &  $\bf{5.73}$ &   55\\
5 & $\bf{2.46}$ &  $\bf{3.05}$ &  $\bf{5.06}$ &  $\bf{5.73}$ &   54\\
6 & $\bf{2.56}$ &         2.70 &         3.64 &  $\bf{5.66}$ &   56 \\
7 & $\bf{2.43}$ &  $\bf{2.77}$ &         4.31 &  $\bf{5.67}$ &   55\\
8 & $\bf{2.68}$ &  $\bf{3.04}$ &         4.17 &  $\bf{5.32}$ &   44\\
9 & $\bf{2.46}$ &  $\bf{2.85}$ &         4.37 &  $\bf{5.13}$ &   39 \\

10 &$\bf{2.80}$ &  $\bf{3.50}$ &  $\bf{4.84}$ &  $\bf{6.07}$ &   55\\
\hline
\\
\end{tabular}
\end{threeparttable}
\end{table*}



We then ranked the models for which we computed the monochromatic images based on their reduced $\chi^2$ ($\chi^2_\mathrm{red}$) of both the photometry and the interferometry of the different bands. 
The model with the lowest $\chi^2_\mathrm{red}$ in each category got the lowest rank. 
The resulting five ranks per model were then added to get a set of models that perform well.

As a second selection step, we selected the 15\% models with the lowest ranks.
We subsequently ray-traced the images of these models at six wavelength channels per band, resulting in chromatic images.
For the $H$ band, we calculated the images at the wavelengths of the six spectral channels of PIONIER.
For the other bands, the wavelengths of the different images are equally distributed over the wavelength range of the bands.
The images were subsequently combined and saved as an image cube. 
Model visibilities were calculated at the \modiflet{\textit{uv}} coordinates of the data by linearly interpolating between the wavelengths. 

The models were then assessed based on their performances on the photometric fluxes, star-to-total flux ratios, hlr, and the over-resolved flux component to have metrics on the disc geometry and the radial structure.
For the over-resolved component, we took the value of the visibility at the shortest baseline(s) of the data and compared it to the predicted value of the model at the same baseline. 
Reported values are at baselines of $B = 4.31$ M$\lambda$ in $H$ and $B = 4.55$ M$\lambda$ in $K$. 
In the $L$ band we took into account the set of short baselines $B <0.6$ M$\lambda$ from the shortest baseline ($B = 2.35$ M$\lambda$) to have a fair comparison between the model and the data by taking into account the uncertainties on the $V^2$.
We took this larger range since there are data points at $V^2 = 0.46$ and at $V^2 = 0.57$ with overlapping uncertainties at the short baselines with an average of $V^2 = 0.51$.
In the $N$ band, the correlated flux for the shortest baselines ($< 0.8 $M$\lambda$) are between $F_\mathrm{corr} \sim 160$ Jy and $F_\mathrm{corr} \sim 220$ Jy.
Reported values for are at $B = 2.47$ M$\lambda$ and $B = 0.67$ M$\lambda$ in $L$ and $N$, respectively.
The confidence levels for which we consider our models predicting sufficient over-resolved fluxes are chosen to be 20\% in the near-IR bands and 40\% in the mid-IR bands, proportional to the measurement uncertainties.

Similarly, we set the confidence levels for which we consider that our models predict a large enough radial extent: 15\% for the near-IR bands and 20\% in the mid-IR bands.
As noted in Sect.\,\ref{section:strategy-exploration}, we took the radial extents as constrained by geometric models in \citet{Corporaal_2021}, which are reported at the central wavelength of the four bands.
To compare the model hlr with these results, we calculated the hlr of the image at these central wavelengths.

The stellar flux contribution with respect to the total, $F_s/F_\mathrm{rest}$, was calculated from the central wavelength image of the $H$ band, as the star is contributing most significantly in this band, with a contribution $\sim 62\%$ to the total flux at 1.65\,$\mu$m \citep{Corporaal_2021}.
The total contribution at this wavelength is coming from the star, the disc, and an over-resolved component.
The total flux emerging from the star, $F_s$ was thus compared with that emerging from the regions farther than the star, $F_\mathrm{rest}$.
The confidence levels for $F_s/F_\mathrm{rest}$ are set to 10\%, proportional to the uncertainties of the geometric models.

We checked whether our models are compliant with the four photometric bands, the stellar contribution as seen in the $H$-band interferometry, the four hlr, the four over-resolved flux components, and the stellar contribution with respect to the total flux in the confidence levels we outlined above.
We summed the amount of times a model complies with these levels.
Each model can get a maximum score of 13.
As a third selection step, we selected the ten models with the highest score and ranked the models with the same score against each other based on their photometric and interferometric $\chi^2_\mathrm{red}$.

\begin{table*}[t]

\caption{$\chi^2_\mathrm{red}$ and \modiflet{short} baselines of the family of best-fit models.}
\label{Table_visibilities_10best}
\centering
\begin{threeparttable}
\begin{tabular}{lccccccccccccccc}
 \\

\hline
Model & $F_s/F_\mathrm{rest}$ & $\chi^2_\mathrm{red, H}$ &       $\chi^2_\mathrm{red, K}$ &       $\chi^2_\mathrm{red, L}$ &     $\chi^2_\mathrm{red, N}$ &  $V^2_\mathrm{small_H}$ &  $V^2_\mathrm{small_K}$ &  $V^2_\mathrm{small_L}$ &  $F_\mathrm{corr, \mathrm{small_N}}$\\ 
& & & & & && & & (Jy) \\

\hline
data &1.65 &-&-& -& - &0.65 & 0.43 & 0.51 & 208  \\
1 & \bf{1.62} & 20 &    313 &    251 &   41 &  $\bf{0.77}$ &  $\bf{0.51}$ &  $\bf{0.67}$ &  $\bf{287}$\\
2 & \bf{1.58} & 12 &    265 &    149 &   29 &  $\bf{0.75}$ &  $\bf{0.50}$ &  $\bf{0.70}$ &  $\bf{283}$\\
3 & 1.45 & 38 &    317 &    189 &   34 &  $\bf{0.74}$ &  $\bf{0.49}$ &  $\bf{0.68}$ &  $\bf{270}$ \\
4 & \bf{1.74} & 21 &    272 &     81 &   28 &  $\bf{0.76}$ &  $\bf{0.51}$ &  $\bf{0.63}$ &  $\bf{287}$\\
5 & \bf{1.70}&21 &    338 &    187 &   22 &         0.81 &         0.53 &  $\bf{0.66}$ &  $\bf{273}$\\
6 & \bf{1.66} & 6 &    242 &    122 &   17 &  $\bf{0.73}$ &  $\bf{0.50}$ &  $\bf{0.62}$ &  $\bf{250}$\\
7 & \bf{1.82} & 15 &    271 &    120 &   24 &         0.79 &  $\bf{0.52}$ &  $\bf{0.64}$ &         301\\
8 & 1.34 & 30 &    301 &    144 &   21 &  $\bf{0.74}$ &  $\bf{0.51}$ &  $\bf{0.67}$ &  $\bf{240}$\\
9 & \bf{1.75} & 17 &    308 &    269 &   32 &  $\bf{0.78}$ &         0.53 &  $\bf{0.67}$ &  $\bf{276}$\\

10 & 1.38 & 57 &    354 &    117 &   11 &  $\bf{0.77}$ &         0.52 &  $\bf{0.65}$ &  $\bf{233}$\\
  
\hline
\\
\end{tabular}
\tablefoot{Reported $\chi^2_\mathrm{red}$ are calculated from the chromatic images considering the full interferometric dataset.}
\end{threeparttable}
\end{table*}

\subsection{Results}
\label{section:results-application}
\subsubsection{The parameters after each selection step}
The results of the three selection steps per parameter are shown in Fig. \ref{fig:selection}.
The total number of models are equally distributed between all parameters.
From the inspection of the SEDs, we \modiflet{find} 329 (28\% of the total) models display SEDs satisfying the criteria defined in Sect.\,\ref{section:strategy-application}. 
From this first selection we can deduce two general parameter preferences.
First, both a high scale height at the inner rim ($h_0 = 1.87$\,au), or a high degree of disc flaring ($\beta = 1.4$) show an overprediction of the $N$ band and to a lesser extent also an overprediction in the $L$-band flux.
Instead, models with a $h_0 = 0.93$\,au and a $\beta=1.2$ are favourable with 53\% and 62\% of the models after the first selection, respectively.
Second, grain size distributions with $q = 2.75$ or $q = 3.0$ are preferred over larger values of $q$, pointing towards the presence of larger grains.

After the second selection, 55 models are left. 
The IR interferometric observations show strong constraints in four of the parameters.
The second selection step shows similar contributions as the first selection step in $\beta$ with a strong preference for $\beta= 1.2$.
Moreover, there is a clear preference for $h_0 =1.40$\,au.
There are strong differences within the wavelength bands. 
The near-IR interferometric data are sensitive to $h_0$ and strongly prefer $h_0 = 1.40$ au.
The $L$-band data show a preference for $h_0=1.40$ au but values of $h_0=0.93$\,au are not ruled out.
The $N$-band data show, however, a clear preference for $h_0 = 0.93$\,au,  resulting from the influence on the amplitude of the second lobe in the interferometry.
Models with higher $h_0$ show higher amplitudes, which resides above the data.
Third, the IR interferometric observations clearly prefer small dust masses.
Dust masses of $1.0 \times 10^{-3}$ M$_\odot$ are highly preferred in 62\% of the models, with a decreasing slope with increasing dust mass.
Fourth, grain size distributions with $q=2.75$ are preferred while none of the models have $q = 3.25$. 

The ten models that remain after the third selection step \modiflet{show} a strong preference for $\beta = 1.2$.
90\% of the models after this selection have $h_0 =1.40$\,au while all models with $h_0 = 0.93$\,au are ruled out.
A scale height at the inner rim larger than for our reference model is thus preferred.
The distribution of dust masses and the grain size distribution power law remain similar to the second selection.
While the first and second selection \modiflet{do} not prefer models with or models without Fe, the third selection shows a strong preferences for models without metallic iron.
The distributions of $\alpha$ and $r_\mathrm{mid}$ after the three selection steps do not show a strong preference for certain values of these parameters.


\subsubsection{The family of best-fit models}
\begin{figure*}
\centering
  \includegraphics[width=17cm]{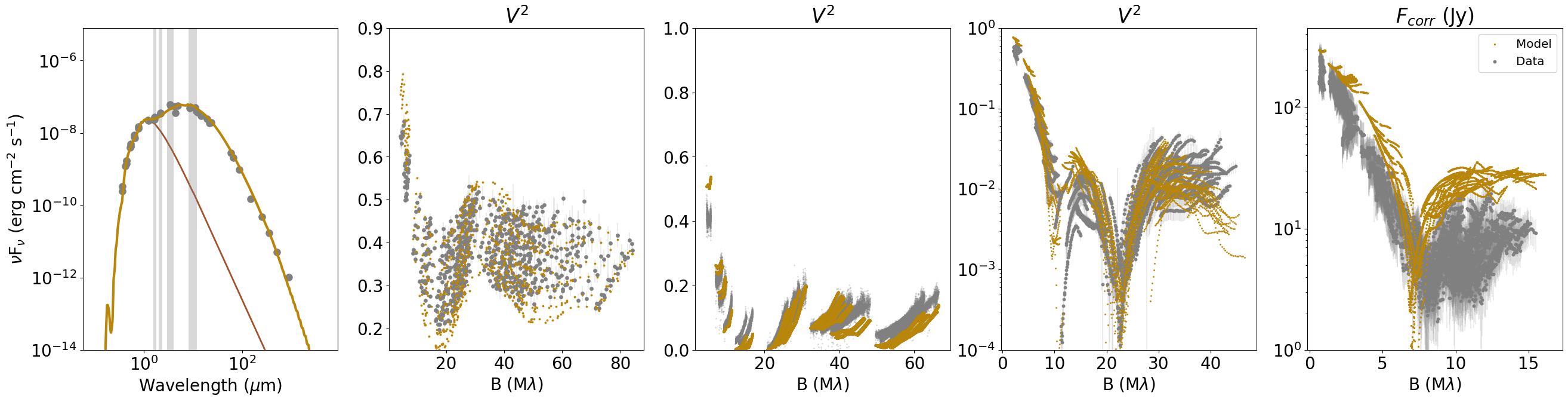}
  \caption{SED and visibility curves as in Fig. \ref{fig:SEDVIS_reference_model} but for Model 1 from our family of best-fit models.
 }
  \label{fig:SEDandVIS_model1}
\end{figure*}

\begin{figure}
\centering
  \resizebox{\hsize}{!}{\includegraphics[width=12cm]{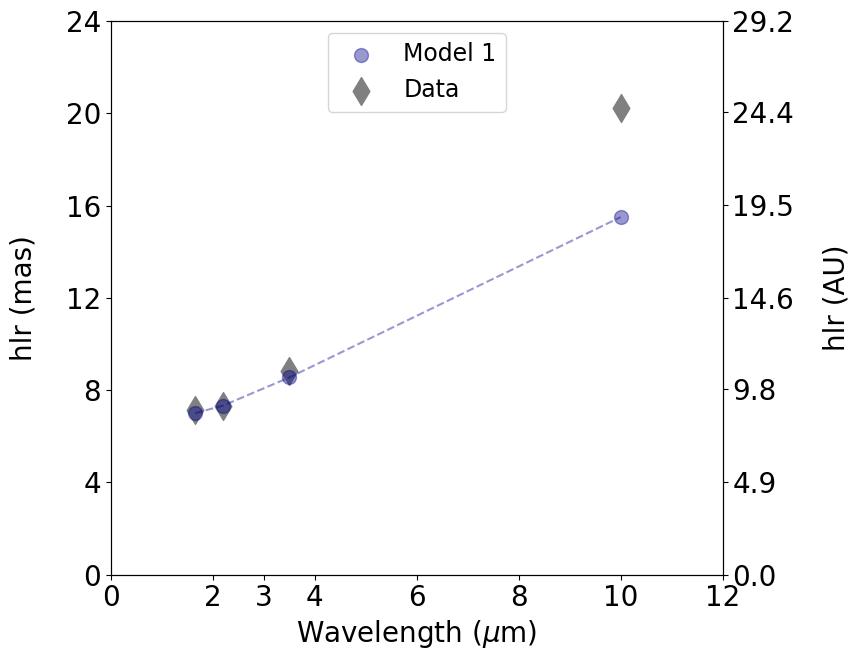}}
  \caption{Half-light radii of Model 1 compared to those from the geometric models of \citet{Corporaal_2021} at the central wavelengths of the $H$, $K$, $L$, and $N$ bands.
 }
  \label{fig:best_model_hlr}
\end{figure}

\begin{figure*}
\centering
\includegraphics[width=17cm]{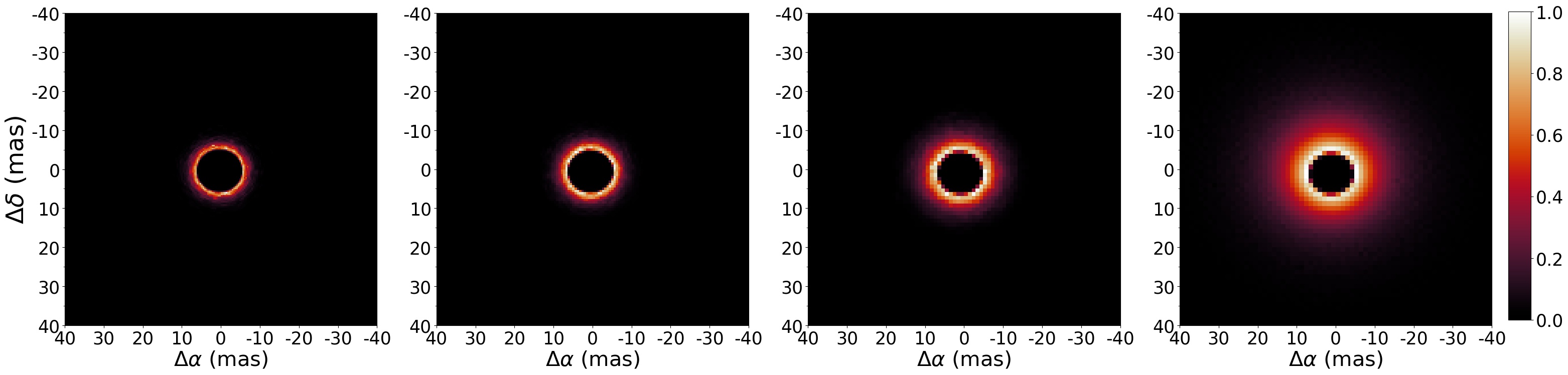}

  \caption{Circumbinary disc images of Model 1 at the central wavelengths of the (from left to right) $H$, $K$, $L$, and $N$ bands.
The fluxes of central stars are removed from the image to unveil the disc structures.
The images are normalised to the total flux of each image.
 }
  \label{fig:best_model_images}
\end{figure*}

The models resulting from the third selection step are our family of best-fit models.
The parameters of these models are displayed in Table \ref{Table:models_10best}.
The performance of these models on \modiflet{the} photometry is displayed in Table \ref{Table:models_SED_10best}.
The stellar contribution, the visibility $\chi^2_\mathrm{red}$ of the different bands resulting from the chromatic images, and the values of the visibility at the smallest baselines of these models are given in Table \ref{Table_visibilities_10best}.
The hlr of these models are given in Table \ref{Table:radialstructure_best10}.
Values represented in bold in Tables \ref{Table:models_SED_10best}-\ref{Table:radialstructure_best10} are within our set confidence levels and are considered to be good enough within our framework.
The score, sc, in Table \ref{Table:models_10best} corresponds to the addition of the number of such bold representations in the tables.
The SED and IR interferometric visibility curves of Model 1 are shown in Fig. \ref{fig:SEDandVIS_model1}.
The performance of Model 1 on the wavelength-dependent radial extents is shown in Fig. \ref{fig:best_model_hlr}. The images of the $H$, $K$, $L$, and $N$ bands of Model 1 are shown in Fig. \ref{fig:best_model_images}.

Overall, our resulting family of models is able to explain many features of our dataset.
Model 1 performs best in terms of the photometry, the over-resolved fluxes and the radial extents and scores 12 out of a maximum of 13.
It is able to reproduce well the photometry, the over-resolved flux component, the stellar contribution, and the radial extent in the $H$, $K$, and $L$ bands. 
It only cannot reproduce the $N$-band radial extent within our set limits. 
Models 4, 5, and 7 are able to reach these levels in the $N$-band radial extent. 
Models 2-5, have a total score of 11 and Models 6-10 have a score of 10.

Models 1-4, 6, and 8 reproduce the over-resolved flux component at short baselines within our set limits. 
The models are thus able to explain at least part of the over-resolved flux.
However, there is a systematic underprediction of the over-resolved flux component compared to the data in all the models (see Sect.\,\ref{section:Discussion_extended} for a discussion).

The photometric flux is well reproduced in all bands for Models 1, 3, 5, and 10.
This shows that the disc geometry is able to reproduce the fluxes in the SED.
The models with metallic Fe, Model 5 and 10, come closer to the $K$- and $L$-band fluxes than all other models, as expected from the parameter study (Sect.\,\ref{section:parameterstudy}).
These models lack, however, over-resolved flux, most notable for Model 5 in the $H$ band.

Model 6 performs best in the IR interferometric $\chi^2_\mathrm{red}$. 
In the $H$ band, the model is the highest ranked in the second and third selection and in the $K$ and $L$ bands it is ranked in the top 10\%, but in the $N$ band it is in the bottom 50\%. 
This is due to the larger $h_0$, which is preferable for $H$, $K$, and $L$ but not for $N$.

\begin{table}
\caption{Half-light-radii of the family of best-fit models.}
\label{Table:radialstructure_best10}
\centering
\begin{threeparttable}
\begin{tabular}{lccccccccccc}
 \\
\hline
Model & hlr$_\mathrm{H}$ & hlr$_\mathrm{K}$ & hlr$_\mathrm{L}$ & hlr$_\mathrm{N}$\\
& (mas) & (mas) & (mas) & (mas) \\

\hline
geometric model& 7.12 & 7.31 & 8.85 & 20.22 \\
1 &$\bf{7.00}$ &  $\bf{7.33}$ &  $\bf{8.55}$ &         15.50\\
2 & $\bf{7.09}$ &  $\bf{7.38}$ &  $\bf{8.65}$ &         16.00\\
3 & $\bf{7.24}$ &  $\bf{7.48}$ &  $\bf{8.65}$ &         14.50\\
4 & $\bf{7.00}$ &  $\bf{7.33}$ &  $\bf{8.65}$ &  $\bf{17.00}$\\
5 & $\bf{6.87}$ &  $\bf{7.24}$ &  $\bf{8.55}$ &  $\bf{17.00}$\\
6 & $\bf{7.47}$ &  $\bf{7.62}$ &  $\bf{9.09}$ &         15.90\\
7 & $\bf{7.18}$ &  $\bf{7.48}$ &  $\bf{8.99}$ &  $\bf{16.40}$\\
8 & $\bf{7.00}$ &  $\bf{7.24}$ &  $\bf{8.42}$ &         14.50\\
9 & $\bf{7.00}$ &  $\bf{7.33}$ &  $\bf{8.55}$ &         15.00\\

10 & $\bf{6.78}$ &  $\bf{7.05}$ &  $\bf{8.32}$ &         16.10\\

\hline
\end{tabular}
\tablefoot{Reported hlr are at the central wavelengths of the interferometric bands.}
\end{threeparttable}
\end{table}

\begin{figure}

 \resizebox{\hsize}{!}{\includegraphics[width=12cm]{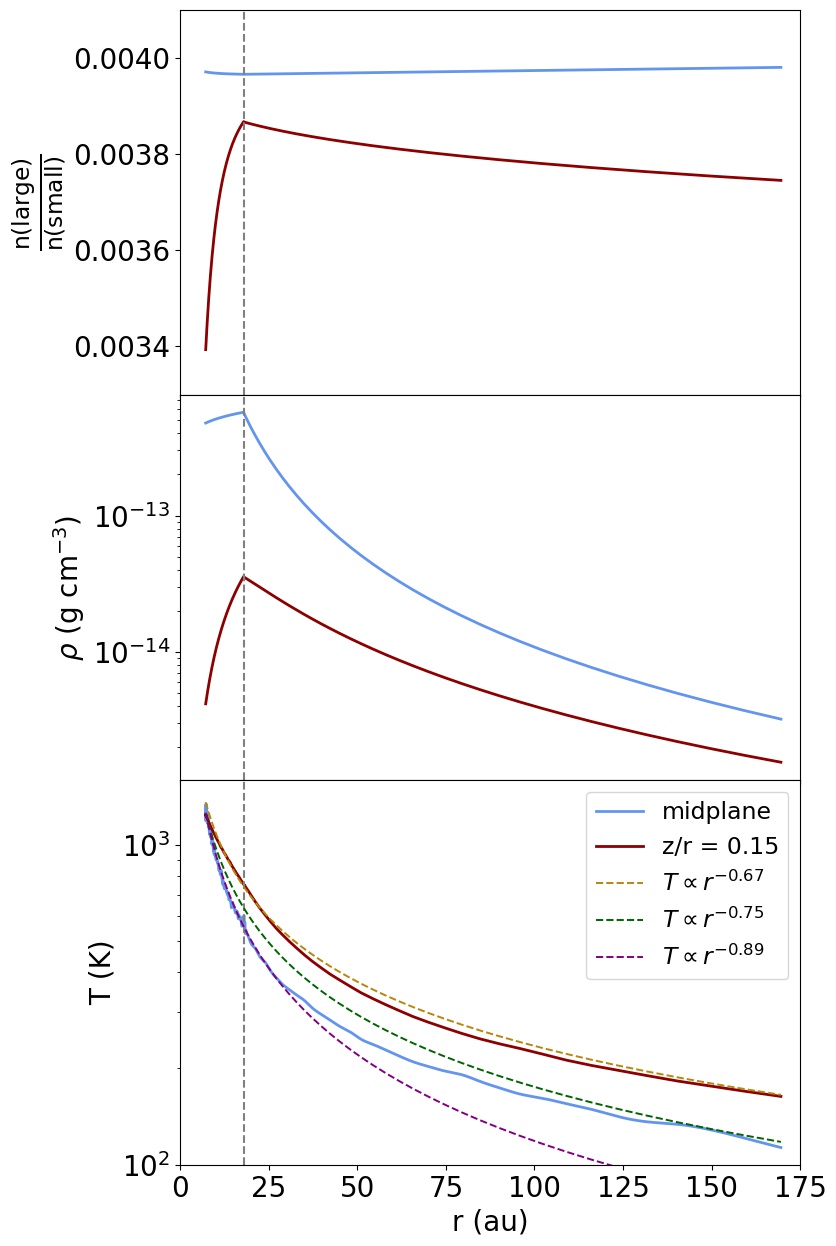}}
 \centering
  \caption{Cuts of the dust distribution, dust density, and dust temperature of Model 1 at the midplane and \modif{at $z/r =0.15$}. 
 The dust distribution is quantified as a ratio of the number of large grains ($>1 \mu$m) to small grains ($<1 \mu$m) and is calculated as a number density per unit mass. 
 The vertical dashed line indicates the location of $r_\mathrm{mid}$.
 \modif{The dashed yellow, green, and purple lines in the bottom plot indicate temperature profiles with power-law indexes of -0.67, -0.75, and -0.89, respectively.}
 }
  \label{fig:best_model_cuts}
\end{figure}


\section{Discussion}
\label{section:discussion}
In this work we show that we can find a family of physical models that \modiflet{reproduces} the extensive dataset covering both photometric and multi-wavelength IR interferometric observations from the $H$ to the $N$ band with a parameterised disc model, originally designed for protoplanetary discs around young stars.
Here we discuss the implications of the results of Sect.\,\ref{section:results-application}.
We discuss the constrained disc parameters and its implications for the disc structure (Sect.\,\ref{section:discstructure}), the origin of the missing over-resolved flux (Sect.\,\ref{section:Discussion_extended}), the difference in preferred structural disc parameters between the $H$, $K$, and $L$ band and the $N$-band interferometric data (Sect.\,\ref{sect:Nbanddisconnection}), the comparison to protoplanetary discs around young stars (Sect.\,\ref{section:comparison_YSOs}), and finally some future prospects (Sect.\,\ref{section:futureprospects}).

\subsection{Constrained disc parameters and disc structure}
\label{section:discstructure}
This disc structure constrains the physical processes in the disc. Our results show that five out of the seven disc parameters can be well constrained with the combination of photometric data and IR interferometric data in four bands (see Fig.\,\ref{fig:selection}).
The following five parameters are constrained: the two disc flaring parameters $\beta$ and $h_0$, the dust mass $m_d$, the index of the grain size distribution, $q$, and the presence or absence of metallic iron as an additional opacity source.

The scale height of the disc is well constrained, with a strong preference for a flaring index $\beta = 1.2$ and a scale height at the inner rim of $h_0 = 1.40$\,au, \modif{indicating that the disc is more vertically extended than expected from the two-dimensional hydrostatic equilibrium model of \citet{Kluska_2018} (i.e. our reference model, $h_0=0.99$\,au), starting at the disc inner rim.}
Moreover, a value of $q = 2.75$ is strongly preferred over larger values of $q$.
This indicates that large grains are present in the inner disc regions, where the IR interferometric observations are sensitive (see also Sect.\,\ref{sect:Nbanddisconnection}).
The IR interferometry also prefers lower dust masses of the order of $m_\mathrm{d} = 1.0 \times 10^{-3}$\,M$_\odot$.
This translates to dust densities of the order of $2.1 - 4.8 \times 10^{-13} \mathrm{g/cm}^3$ at the inner rim for for models with metallic iron and models without metallic iron, respectively. 
Dust masses above $\sim 2.0 \times 10^{-3}$\,M$_\odot$ are ruled out at the assumed distance of the system. 
The final selection also prefers discs without metallic Fe.
The opacity of metallic iron affects strongly the dust density at the inner rim, which we can constrain with our IR interferometric observations. The factor of two difference between this density with the addition of metallic iron and without this addition points towards a constrained inner rim dust density of $4.8 \times 10^{-13} \mathrm{g/cm}^3$.
Finally, the degree of dust settling, $\alpha$, and the turn-over radius, $r_\mathrm{mid}$, are not well constrained with our dataset.


To illustrate the structure of the disc model, we present the dust grain distribution, the dust density, and dust temperature of Model 1 at two values of the vertical scale height, $z/r$, in Fig.~\ref{fig:best_model_cuts}.
We consider the fraction between large and small dust grains within the disc a good measure for the dust grain distribution throughout a disc.

As expected, the number of large grains is higher in the midplane than at larger scale heights in the disc, as larger grains \modif{are settled towards the midplane.}
\modif{The dust density decreases vertically, as expected for these discs, as the small grains populate the surface layers and the large grains settle towards the midplane.
The dust temperature increases vertically as the grains populating the midplane are more shielded from direct stellar radiation.}
The radial kink in the dust density distribution at the turn-over radius results from the changing surface density distribution in our two-zone model (Eq.\,\ref{Eq:surfacedensity}).

The radial and vertical dust temperature structure is well constrained.
The dust density profile, however, depends on $r_\mathrm{mid}$, which is not strongly constrained (larger values are slightly preferred) with our photometric and IR interferometric observations, and on the dust mass in the disc.
These are intertwined, as the surface density changes, depending on the turn-over radius, which is also linked to the dust mass in the two zone model as the two zones are well connected. Within our family of best models, the dust density at the inner regions can differ by a factor of up to two for larger values of $r_\mathrm{mid}$.




\begin{figure*}

\centering
     \includegraphics[width=17cm]{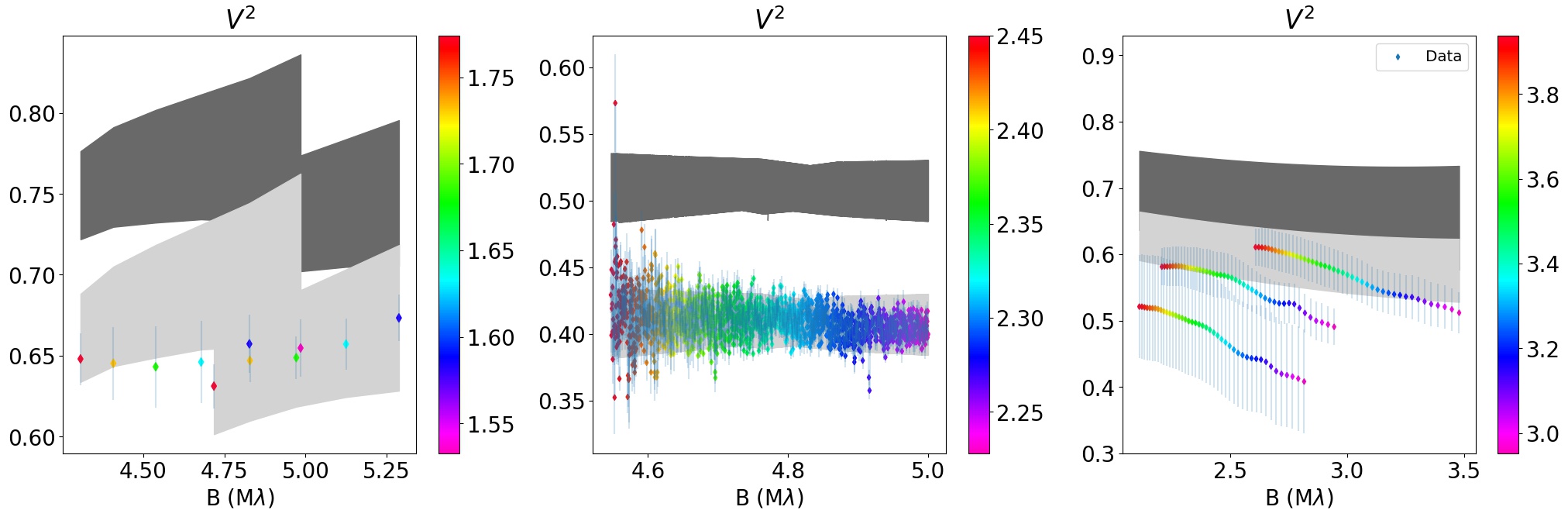}

  \caption{Range of squared visibilities at the shortest baselines of the family of ten best-fit models (shaded dark grey) and of those models with an addition of an over-resolved component with a relative contribution of 10\% and a temperature of 2100\,K (shaded light grey) for the (from left to right) $H$, $K$, and $L$ bands.
  }
  \label{figure:short_baselines_fits}
\end{figure*}

\subsection{The missing over-resolved flux}
\label{section:Discussion_extended}


The excess of over-resolved flux in near-IR VLTI data \modiflet{has been} noticed in previous modelling efforts of post-AGB discs \citep{Hillen_2014, Kluska_2018,Kluska_2019}.
None of the radiative transfer models in this work are able to reproduce the total over-resolved flux, which means that its origin is still unknown. 

We aim to investigate the origin of the over-resolved flux component by reproducing both the photometric fluxes up to 20\,$\mu$m, and the short baselines squared visibilities $V^2_\mathrm{small}$ in the $H$, $K$, and $L$ bands.
The $N$-band data are displayed in correlated flux and are therefore not sensitive to the over-resolved emission.

In Sect.\,\ref{section:results-application} we tested whether the disc geometry (flaring, vertical dust settling) could be the origin of the over-resolved flux.
Our family of best-fit models shows that in the $H$, $K$, and $L$ bands, $\sim$50\% of the over-resolved flux that was missing in our reference model has been captured.
Therefore, the disc geometry can only account for a part of the over-resolved flux component.
Moreover, while our confidence levels of 20\% and 40\% in the near-IR and mid-IR bands, respectively, are reached for all bands in six out of ten models, there is a systematic overestimation of the visibility at \modiflet{short} baselines.
Clearly, some over-resolved flux is still missing.

Similarly to \citet{Kluska_2018}, we modelled the over-resolved emission with a single temperature blackbody normalised to 2.2$\,\mu$m, which is about the central wavelength over the $H$, $K$, and $L$ bands.
We added this component to the interferometry of our family of best-fit models.
We fitted the following two parameters: the relative flux of the over-resolved component with respect to the total flux and its temperature.
We made a grid over these two parameters with temperatures in the range of $T_\mathrm{back}$ from $400$ to $7000$\,K and flux contributions in the range of $F_\mathrm{back}$  $1$ to $10\%$. 
For the photometry, we used a Levenberg-Marquardt fitting routine from the Python package \modiflet{\textsc{lmfit}} and evaluated the median and range of the best-fit values to all the different models in the family of best models.
For the interferometry, we evaluated the over-resolved emission by computing the reduced $\chi^2$ on the shortest baselines of the $H$-, $K$-, and $L$-band interferometric data. We \modiflet{quantified} which combination of parameters reproduces the short baselines visibilities in all bands the best. 
We subsequently selected 5\% of the models with the lowest total reduced $\chi^2$ values in the grid and computed the median temperature of their contributions.


We find a better fit to the photometric fluxes if an additional flux component with a median temperature of 1470\,K and a median relative contribution of $7$\% at 2.2\,$\mu$m is added.
The ranges in best-fit temperatures and relative flux contributions we fitted to the family of best models are 1120-2100\,K and 2-10\%, respectively.
This relative flux component is not very well constrained and depends on the individual model.
For Model 1, it is estimated to $F_\mathrm{back}=7\pm4\%$ in $K$ with $T_\mathrm{back}=1610\pm450$\,K.
This would not significantly decrease its $\chi^2_\mathrm{red,SED}$ from 45 to 44 and increase the photometric fluxes $F_\mathrm{H}$, $F_\mathrm{K}$, $F_\mathrm{L}$, and $F_\mathrm{N}$ by 4\%, 7\%, 8\%, and $3\%$, respectively, which brings them closer to the values of the data in $H$, $K$, and $L$.

The short baselines are fitted better with an additional over-resolved component with a median temperature of $2100$\,K and a contribution of 10\% to the relative flux in $K$, with a range in temperatures of 1400-3600\,K and a relative contribution of 8-10\% within the family of best-fit models.
The impact of this addition on the short baseline interferometric observables is shown in Fig. \ref{figure:short_baselines_fits}. 
The $V^2$ of the family of best-fit models are clearly lying above the data for all bands.
With the added component, the $V^2$ can go down to the $V^2$ of the data within the uncertainties in the $H$ and $K$ bands.
For the $L$ band it reaches within $V^2 \sim 0.6$ while it does not reach within the lower-lying $V^2$ at $B<3.0\mathrm{M}\lambda$.



While photometry does not allow strong constraints to be placed on the extended emission, the interferometric data are clearly better reproduced with the addition of such a component.
Its temperature of 2100\,K (with a range between 1400 and 3600\,K) points towards a mixed origin of photons that may come from both the star (T$_\mathrm{eff}$ of 7250\,K) and the hot inner disc rim (T$_\mathrm{eff}$ of $\sim$1250\,K). 

The exact location and geometry of the component are, however, not constrained by our data and it is, thus, difficult to claim what would be the disc geometry or the morphology of an additional component that could reproduce our data.

Several other possibilities have been discussed in \citet{Kluska_2018} and should be tested in future studies. 
One such possibility is that the emission originates from the jet launched from the accretion disc around the secondary star.
These jets are commonly observed in post-AGB binaries and are detected in absorption during orbital phases when the companion is in front of the post-AGB primary \citep[and references therein]{Bollen_2022}. Such a jet is also detected in IRAS\,08544 (Kluska et al. 2018). It is, however, as yet unknown if also dust grains are ejected in these jets.

\subsection{The N-band disagreement}
\label{sect:Nbanddisconnection}
While we can obtain a good fit the multi-wavelength IR interferometric data and the photometry at the same time, there seems to be a general disagreement with the $N$-band interferometric data.
This is notable in both our second selection and the third selection of Sect.\,\ref{section:application}.
The $N$-band data prefer models with \modiflet{(1)} a scale height at the inner rim of $h_0 =0.93$\,au to lower the amplitude of the second lobe, \modiflet{(2)} the presence of metallic iron, and \modiflet{(3)} a grain size distribution index of $q=3.0$.
Our global selection, however, rules out low $h_0 = 0.93$\,au and disfavours higher values for $q$ and the presence of metallic iron.
This is due to the finding that the $H$-, $K$-, and $L$-band interferometric data do not prefer these values.

This $N$-band disagreement is best illustrated in Model 1.
While it performs best in our total set of criteria, it has the worst value of the $\chi^2$ on the $N$-band interferometry.
Moreover, while the wavelength-dependent radial extent is always reached in the $H$, $K$, and $L$ bands for our ten best models, the $N$-band extent is only reached in three out of ten models, in which there is still a systematic under prediction of the extent (see also Fig. \ref{fig:best_model_hlr}).
Here we want to comprehend what conditions cause this disagreement between the $H$, $K$, and $L$ bands, which probe the very inner disc regions close to the inner rim, and the $N$ band, which probes deeper into the disc.

The best model for the $N$-band interferometry, independent of the photometric or other interferometric data, has a small scale-height at the inner rim, $h_0=0.93$\,au, an $\alpha = 0.01$, a dust mass of $m_d= 1.0$\,M$_\odot$, a turn-over radius of $r_\mathrm{mid} = 3.5 \times \mathrm{R_{in}}$, and a dust size distribution index $q=3.0$ with metallic iron mixed with the silicates throughout the disc.
This difference in scale height at the inner rim with respect to the family of best-fit models significantly alters the dust density at different vertical heights, as a disc with a higher $h_0$ has more volume to distribute the particles of the same mass than a disc with a smaller $h_0$.
For this reason, the dust density at the midplane is higher for smaller $h_0$ and the dust temperature is lower.
The systematic higher $q$ in the best ranked models within the $N$ band compared to the other bands indicates that the outer regions contain more small grains compared to the inner disc regions.

For future works, we suggest considering distinctions between the inner and outer disc regions within the dust distribution, the dust species, and/or structural parameters.
This cannot be achieved by changing one parameter as there should be a balance between decreasing the $N$-band photometric flux (as it is systematically over-estimated in the family of best-fit models) and increasing the radial extent in $N$ while the extents in $H$, $K$, and $L$ remain unchanged.

\subsection{Comparison with YSOs}
\label{section:comparison_YSOs}
The family of best-fit models shows that a strong dust settling \modif{and a larger scale height starting at the inner rim are} needed to explain part of the over-resolved flux component as well as the wavelength-dependent radial intensity profile. 
Flaring disc shapes in protoplanetary discs have been confirmed by various direct observations \citep[e.g.][]{Ginski_2016, Pinte_2018}.
Disc flaring in general is thought to be due to the radial internal temperature of the disc decreasing slower than $r^{-1}$ \citep{KenyonHartmann_1987}.
\modif{Two-dimensional geometric modelling in the image plane} of the near-IR and mid-IR interferometric data of IRAS\,08544 confirms that this is also the case for this circumbinary disc around an evolved binary system \citep{Corporaal_2021}. 
In this work, \modif{we also show a temperature decrease with a power of $-0.67$ at the surface layers.
The midplane temperature decreases with a power of $-0.89$ at the inner regions and with $\sim -0.75$ at the outer regions} (see Fig. \ref{fig:best_model_cuts}).

Strong turbulence mixing strengths ($\alpha = 0.1$ and $\alpha =0.01$) are found in our disc modelling to explain the SED characteristics as well as the IR interferometric visibilities.
These are higher than found for most protoplanetary discs around YSOs for which values of $\alpha = 10^{-4}-10^{-2}$ are consistent with observations \citep[e.g.][]{Mulders_2012}.
These authors also show that this turbulent mixing strength depends on the adopted grain size distribution and gas-to-dust ratio.
Higher turbulence parameter values are found for a higher $q$ or a lower gas-to-dust ratio.
We note that $\alpha$ is not well constrained (see Fig. \ref{fig:selection}).
While in $H$, $K$, and $L$ lower values of $q=2.75$ are preferred, higher values are preferred in $N$, indicating that the turbulence parameter could take different values in different disc zones and should be taken into consideration.
In future studies, also differences in the gas-to-dust ratios in different zones of the disc could be taken into account.
Observational these gas properties can be constrained using the \modiflet{Atacama Large Millimeter/submillimeter Array (ALMA)} and the James Webb Space Telescope (JWST).

The dust mass in the disc as constrained from the photometry and the IR interferometry are of the order of $\sim 1 \times 10^{-3}$\,M$_\odot$, which translates in a total mass of $\sim 0.1$\,M$_\odot$ for the assumed gas-to-dust ratio. 
From an analysis on different star forming regions, \citet{Pascucci_2016} found that the total protoplanetary disc masses range from about $0.1$-$1.0$\,M$_\odot$ and that there is an increasing trend with stellar mass.
From this, with our combined stellar mass of 2.4\,M$_\odot$, we could expect dust masses of $\sim 1 \times 10^{-3}$\,M$_\odot$, which is aligned with our deduced dust masses.
The combined stellar mass also makes them close to the average stellar mass of Herbig stars, which have an average dust mass of $4 \times 10^{-4}$ up to $0.1$\,M$_\odot$ \citep{Stapper_2022}, with a generally lower dust mass for non-structured full discs \citep{vanderMarel_2021}.
The inferred dust masses for the disc around IRAS\,08544 are therefore very similar to the higher mass end of the Herbig stars and thus similar to the of the ones found in protoplanetary discs around YSOs.




\subsection{Future prospects}
\label{section:futureprospects}
A limitation of our refined radiative transfer models presented here is that the system is assumed to be axisymmetric.
Image reconstructions of the inner disc regions and the closure phase signal, however, reveal that the disc inner rim shows significant azimuthal variations \citep{Hillen_2016}. 
In future modelling, the binary nature of the system needs to be taken into account to include the asymmetric illumination of the post-AGB star on the disc inner rim during orbital motion.
The orbital phase of this asymmetric illumination will be important to constrain the inner rim structure.
To fully constrain the physical processes in the disc, both the inner disc region and throughout the full disc, we need to combine high-angular-resolution instruments at different spatial scales and at different orbital phases to obtain the full three-dimensional orbit.


\section{Conclusion}
\label{section:conclusion}
We present radiative transfer models for the circumbinary disc around the post-AGB binary system IRAS\,08544 that are able to reproduce an extensive dataset of photometric data and IR interferometric visibilities from the $H$ to the $N$ band. 
The parameterised passive disc models characterise the structure of the inner regions of the circumbinary disc well.

We first explored the impact of the individual parameters on the observables and selected the seven parameters (i.e. turbulence mixing strength, the two scale-height structural parameters, the dust mass, the dust size distribution power-law  index, the turn-over radius of the surface density, and metallic iron) that have substantial effects on the geometry and the radial structure of the disc.
These seven parameters were used for a thorough grid search to isolate models that reproduce the photometric and interferometric dataset.
With combinations of these disc parameters, we aimed to reproduce the characteristics of the dataset with a focus on the over-resolved flux component, the radial structure, and the photometric features.

Our radiative transfer modelling can indeed reproduce the IR visibility data of the disc as well as the photometric features.
We find that the disc has a flared geometry \modif{with a vertical extension starting at the disc inner rim.} 
The $H$-, $K$-, and $L$-band interferometric data, which are sensitive to the inner parts of the disc, prefer a low grain-size distribution power-law  index or, similarly, more large grains, while the $N$-band data prefer higher values.
This points towards evidence for dust grain growth in the inner disc regions.
The dust masses of our preferred models are very similar to dust masses inferred from observations of protoplanetary discs around young stars.

The disc geometry can explain $\sim50$\% of the over-resolved flux component, indicating that scattering partially explains this component, but there is still some over-resolved flux missing.
This missing component has a temperature of the order of $1400-3600$\,K, pointing towards thermal photons emitted from the inner rim that are scattered farther away in the system (from a complex disc morphology or the presence of a halo).
The location and geometry of this component is not constrained with our dataset.

We conclude that the circumbinary disc around IRAS\,08544 is in many ways very similar to the discs around YSOs. 
This bright object is therefore a very good candidate to search for disc asymmetries and eventual signs of macroscopic structure formation around evolved systems.
A full analysis of the asymmetries as well as the gas and dust distributions will be the subject of future research. 
A complete model of the disc with an additional extended dusty component, including the effect of the asymmetric illumination of the binary stars and an adequate connection between the inner and outer disc, awaits systematic studies in which high-angular-resolution data at different spatial scales are combined using observational constraints from the VLTI, \modiflet{the Spectro-Polarimetric High-contrast Exoplanet
REsearch instrument on the VLT}, and ALMA.
Moreover, JWST will allow us to look for the gaseous species inside the dust sublimation radius.





\begin{acknowledgements}
\modif{We thank the referee for their fast and constructive comments and suggestions that substantially improved the clarity of the paper.}
A.C. and H.V.W. acknowledge support from FWO under contract G097619N. J.K. acknowledges support from FWO under the senior postdoctoral fellowship (1281121N). 
DK acknowledges the support of the Australian
Research Council (ARC) Discovery Early Career Research Award (DECRA) grant (DE190100813).
DK is supported in part by the Australian Research
Council Centre of Excellence for All Sky
Astrophysics in 3 Dimensions (ASTRO 3D),
through project number CE170100013.
This work has made use of data from the European Space Agency (ESA) mission Gaia (https://www.cosmos.esa.int/gaia), processed by the Gaia Data Processing and Analysis Consortium (DPAC, https://www.cosmos.esa.int/web/gaia/dpac/consortium). Funding for the DPAC has been provided by national institutions, in particular the institutions participating in the Gaia Multilateral Agreement.
\end{acknowledgements}

\begingroup
\bibliographystyle{aa} 
\bibliography{biblio.bib} 
\let\clearpage\relax
\begin{appendix}

\section{Parametric study}
Similarly to Figs. \ref{fig:parameters_SEDVIS_mostimpacting} and \ref{fig:hlr_5impacting_params}, Figs. \ref{fig:parameters_SEDandVIS_rest} and \ref{fig:hlr_rest_params} show variations in the parameters but for those that have no significant impact on the SED, visibilities, and hlr.

\begin{figure*}
\centering
   \includegraphics[width=17cm]{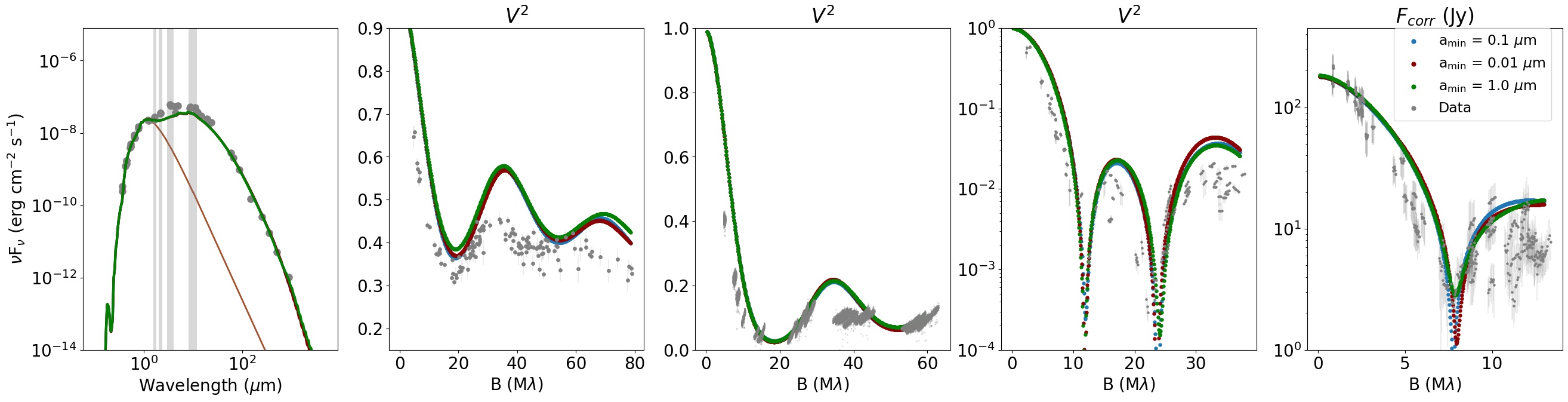}
   
   \includegraphics[width=17cm]{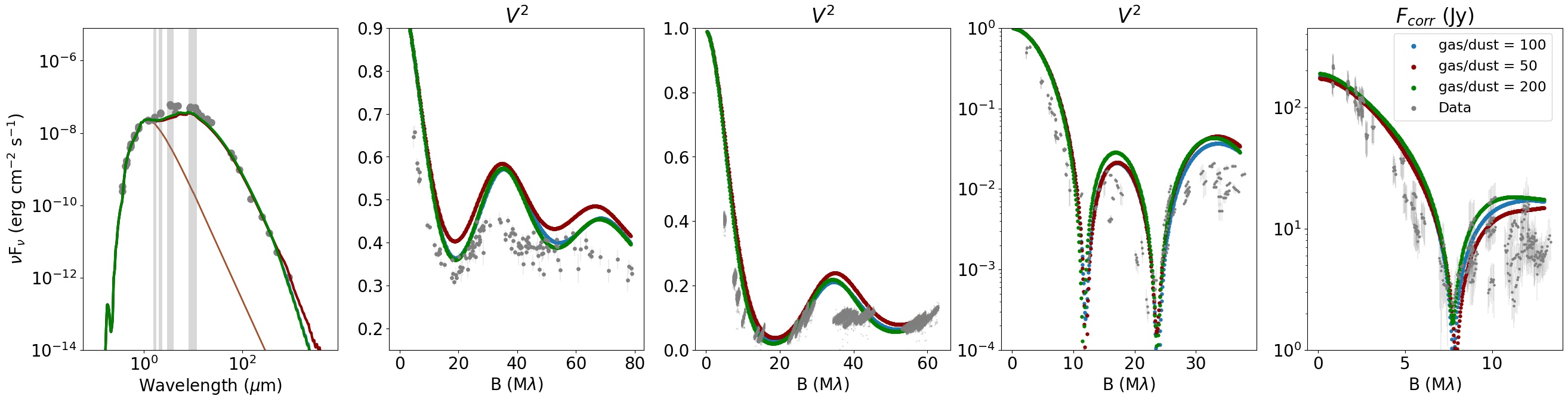}
 
  \includegraphics[width=17cm]{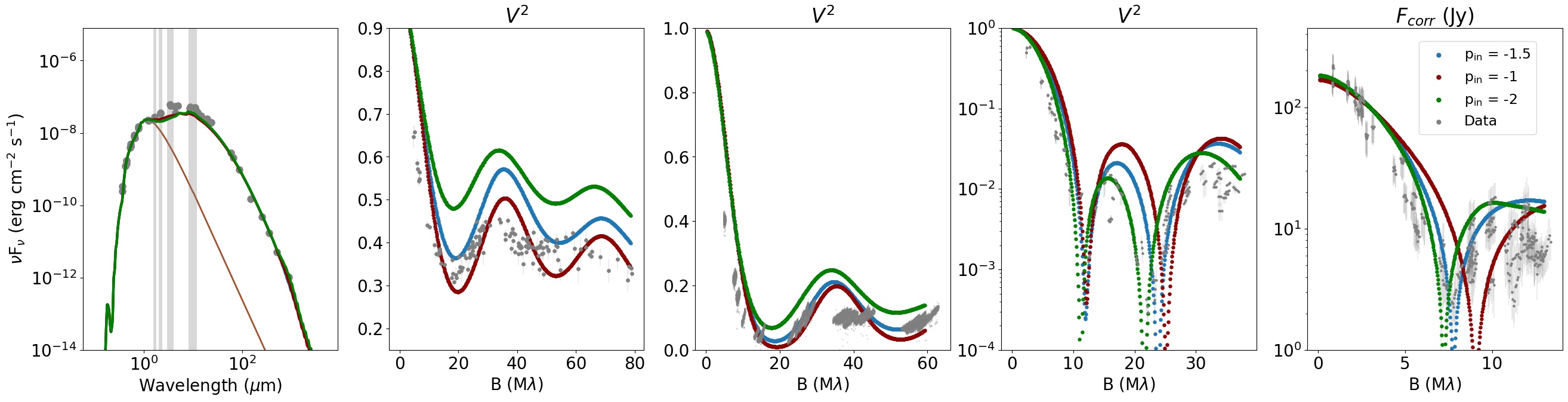}
   
     
 \caption{Variations in\  the parameter study for the parameters that do not significantly impact the observables. 
 From top to bottom: variations in $a_\mathrm{min}$, the gas-to-dust ratio, and $p_\mathrm{in}$}
 \label{fig:parameters_SEDandVIS_rest}
\end{figure*}

\begin{figure*}
\centering
  \begin{minipage}{0.33\textwidth}
  \includegraphics[width=\textwidth,width=1.0
  \textwidth]{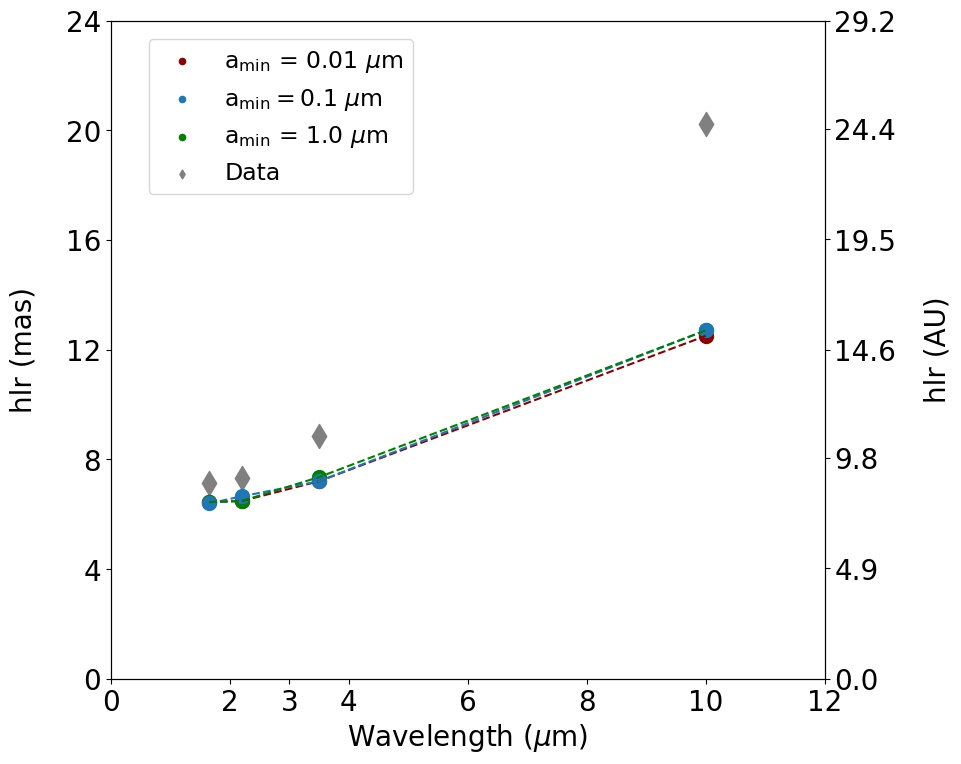}
  \end{minipage}
  \begin{minipage}{0.33\textwidth}
  \includegraphics[width=\textwidth,width=1.0
  \textwidth]{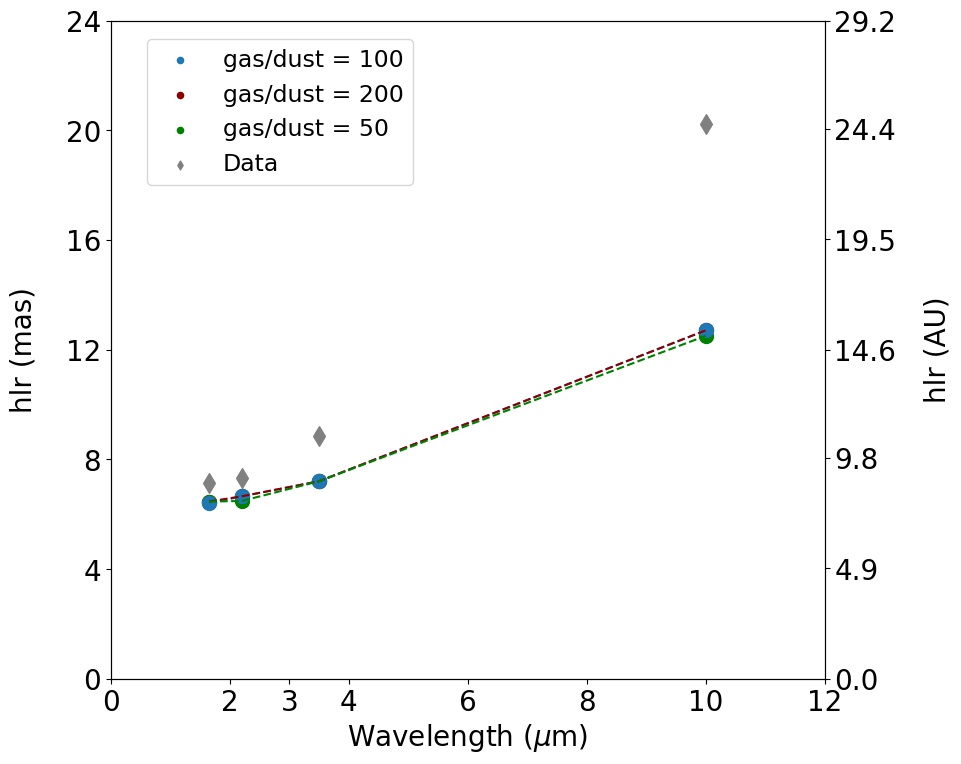}
  \end{minipage}
  \begin{minipage}{0.33\textwidth}
  \includegraphics[width=\textwidth,width=1.0
  \textwidth]{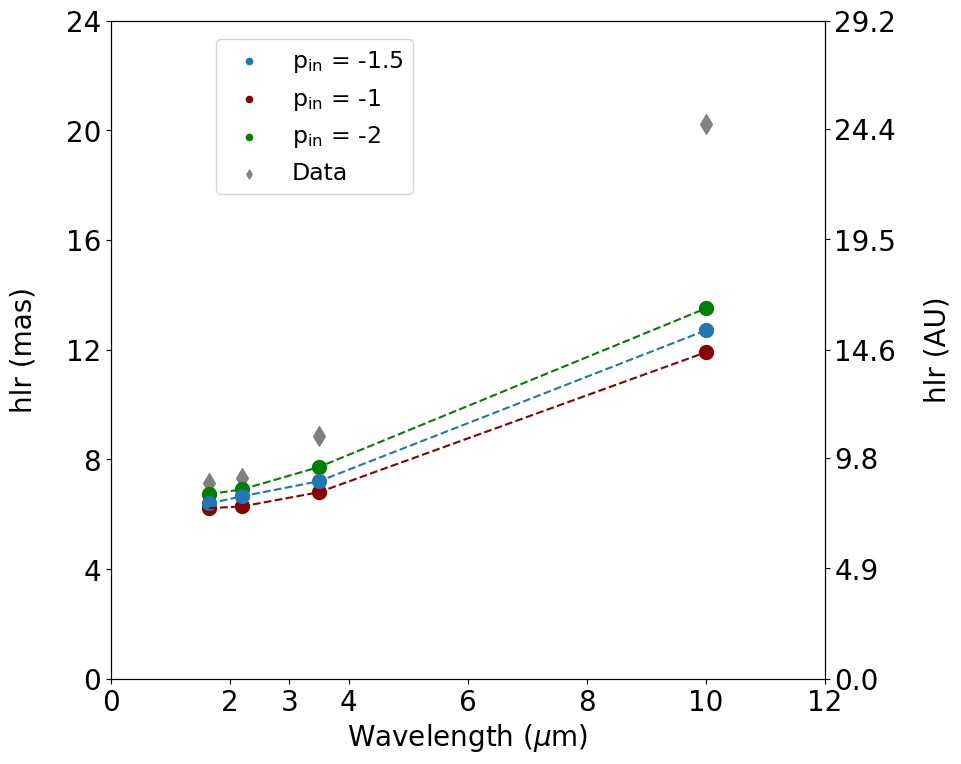}
  \end{minipage}
  

  \caption{Half-light-radius variations in the radial emission as a function of wavelength for the parameters that do not have significant effects on the photometric and/or the interferometric observables. The hlr were calculated at the central wavelengths of the $H$, $K$, $L$, and $N$ bands.
 \textit{Left.} Effect of changing $a_\mathrm{min}$. \textit{Middle}. Effect of changing the gas-to-dust ratio. \textit{Right}. Effect of changing $p_\mathrm{in}$.
 }
  \label{fig:hlr_rest_params}
\end{figure*}
 
 \section{Applied reddening}
 \label{appendix:reddening}
In our modelling, we fitted the interstellar reddening to the photosphere of the post-AGB star. 
The total reddening of the star consists of a circumstellar and an interstellar part.
The circumstellar reddening can change from model to model, as the disc geometry changes with different combinations of parameters within our parameters ranges affecting the reddening properties.
The circumstellar reddening is taken into account in MCMax3D using the extinction law of the interstellar medium as a function of wavelength and the total opacity of the dust grains used in the models.
For each model, the interstellar reddening component is not subject to change substantially, as we look at the disc almost pole-on and no additional structures are placed within the line of sight. 
Literature values of this interstellar reddening component show a large range of values due to the galactic latitude of IRAS\,08544 of 0.3$^\circ$ locating the target close to the galactic plane.
Due to the proximity to the galactic plane and the relatively large distance of $>1$ kpc, interstellar extinction maps do not give reliable results for IRAS\,08544.
The fitted total line-of-sight reddening, $\mathrm{E(B-V)}$, is between 1.36 mag and 1.43 mag for models with an acceptable $\chi^2_\mathrm{red, SED}$ ($<90$). 
The $\mathrm{E(B-V)}$ reaches values of up to 1.47 mag for a few models with an overly flared geometry. 
Our families of best-fit models yield a $\mathrm{E(B-V)} = 1.366-1.38$ mag.
This is in agreement with the fit of \citet{Kluska_2022}, as expected for such a pole-on disc.

\end{appendix}
\endgroup

\end{document}